\documentclass[lettersize,journal]{IEEEtran}
\usepackage{amsmath, amssymb, amsthm, amsfonts}
\usepackage[hidelinks]{hyperref}
\usepackage{graphicx}
\usepackage{booktabs}
\usepackage{bm}
\usepackage{cite}
\usepackage{color}
\usepackage{mathdots}
\usepackage{optidef}
\usepackage{soul}
\usepackage[caption=false,font=normalsize,labelfont=sf,textfont=sf]{subfig}

\usepackage{orcidlink}

\usepackage{textcomp}

\def\BibTeX{{\rm B\kern-.05em{\sc i\kern-.025em b}\kern-.08em
		T\kern-.1667em\lower.7ex\hbox{E}\kern-.125emX}}
\AtBeginDocument{\definecolor{ojcolor}{cmyk}{0.93,0.59,0.15,0.02}}

\DeclareMathOperator{\diag}{\text{diag}}

\newcommand{\unitvec}[1]{\hat{#1}}

\newcommand{\imag}{j}

\newcommand{\conj}[1]{\overline{#1}}
\newcommand{\herm}{\mathsf{H}}

\newcommand{\complex}{\mathbb{C}}
\newcommand{\real}{\mathbb{R}}

\newcommand{\integer}{\mathbb{Z}}

\newcommand{\unitsphere}{\mathbb{S}^2}
\usepackage{tikz}
\usepackage{pgfplots}
\pgfplotsset{compat=newest}
\usetikzlibrary{calc, positioning, backgrounds, arrows, arrows.meta, bending, decorations.text}
\usepgfplotslibrary{groupplots}

\pgfplotsset{
	layers/my layer set/.define layer set={
		bg,
		main,
		foreground
	}{
	},
	set layers=my layer set,
}

\colorlet{labelcolor}{white!25!black}
\colorlet{ticklabelcolor}{white!45!black}
\colorlet{gridcolor}{white!92!black}

\colorlet{legendbordercolor}{white!100!black}

\pgfmathsetmacro{\groupplotsep}{17}
\pgfmathsetmacro{\timeplotheight}{2.8}
\pgfmathsetmacro{\timeplotwidth}{7.9}
\pgfplotsset{time-plot/.style={
		scale only axis,
		legend cell align={left},
		legend style={fill opacity=0, draw opacity=0, text opacity=1, draw=legendbordercolor, font=\scriptsize},
		tick align=inside,
		tick pos=left,
		xmajorgrids,
		ymajorgrids,
		grid style={gridcolor, line width=0.2pt, opacity=0.4},
		ticklabel style = {font=\scriptsize, ticklabelcolor,},
		y label style = {font=\footnotesize, labelcolor, rotate=-90, at=(ticklabel cs:1.1), anchor=west, inner sep=0pt},
		x label style = {font=\footnotesize, labelcolor},
		ytick style={draw=none},
		xtick style={draw=none},
		axis line style={draw=none},
		ylabel shift = -3 pt,
		xlabel near ticks,
}}

\pgfplotsset{position-plot/.style={
		scale only axis,
		axis equal image,
		legend cell align={left},
		legend style={fill opacity=0, draw opacity=1, text opacity=1, at={(0.03,0.97)}, anchor=north west, draw=legendbordercolor, font=\scriptsize},
		tick align=inside,
		grid style={gridcolor, line width=0.2pt, opacity=0.4},
		xmajorgrids,
		ymajorgrids,
		tick style={draw=none},
		ticklabel style = {font=\scriptsize, ticklabelcolor},
		y label style = {font=\footnotesize, labelcolor},
		x label style = {font=\footnotesize, labelcolor},
		xlabel = {x (m)},
		ylabel = {y (m)},
		axis line style = { draw = none },
}}

\pgfplotsset{soundfield-plot/.style={
		axis on top=true,
		scale only axis,
		axis equal image,
		axis line style = { draw = none },
		tick align = inside,
		tick pos=left,
		xtick style={color=black},
		ytick style={color=black},
		ticklabel style = {font=\scriptsize, ticklabelcolor},
		y label style = {font=\footnotesize, labelcolor},
		x label style = {font=\footnotesize, labelcolor},
		xlabel = {x (m)},
		ylabel = {y (m)},
		ylabel shift = -10 pt,
}}

\pgfplotsset{matrix-plot/.style={
		axis on top=true,
		scale only axis,
		axis equal image = true, 
		axis line style = { draw = none },
		tick align = inside,
		tick pos=left,
		xtick style={color=black},
		ytick style={color=black},
		ticklabel style = {font=\scriptsize, ticklabelcolor},
		y label style = {font=\footnotesize, labelcolor},
		x label style = {font=\footnotesize, labelcolor},
}}

\pgfplotsset{spectogram-plot/.style={
		axis on top=true,
		scale only axis,
		axis line style = { draw = none },
		tick align = inside,
		tick pos=left,
		xtick style={color=black},
		ytick style={color=black},
		ticklabel style = {font=\scriptsize, ticklabelcolor},
		y label style = {font=\footnotesize, labelcolor},
		x label style = {font=\footnotesize, labelcolor},
		ylabel shift = -10 pt,
		colorbar style={
			axis equal image = false,
			width=0.2*\pgfkeysvalueof{/pgfplots/parent axis width},
			height=0.8*\pgfkeysvalueof{/pgfplots/parent axis height},
			ylabel shift = 0 pt,
		},
}}

\pgfplotsset{
	micmarker/.style={
		semithick, 
		mark=*, 
		mark size=1.8, 
		only marks,
		draw=black,
		line width=0.5,
		mark options=solid,
}}

\pgfplotsset{
	srcmarker/.style={
		semithick, 
		mark=square*,
		mark size=3,
		only marks,
		draw=black, 
		line width=0.7,
		mark options=solid,
}}

\pgfmathsetmacro{\standardlinewidth}{1.5}
\pgfmathsetmacro{\offamount}{\standardlinewidth*1.7}
\pgfmathsetmacro{\standardlineopacity}{0.9}

\newcommand{\linejoinchoice}{round}

\pgfplotsset{
	line0/.style={
		color0, 
		opacity=\standardlineopacity, 
		line cap=round, 
		line width = \standardlinewidth pt, 
		mark options={solid},
		line join=\linejoinchoice,
}}

\pgfplotsset{
	line1/.style={
		color1, 
		opacity=\standardlineopacity, 
		dash pattern=on 1pt off \offamount pt,
		line cap=round,
		line width = \standardlinewidth pt, 
		mark options={solid},
		line join=\linejoinchoice,
}}

\pgfplotsset{
	line2/.style={
		color2, 
		opacity=\standardlineopacity, 
		dash pattern=on 3.5pt off \offamount pt, 
		line cap=round,
		line width = \standardlinewidth pt, 
		mark options={solid},
		line join=\linejoinchoice,
}}
\pgfplotsset{
	line3/.style={
		color3, 
		opacity=\standardlineopacity, 
		dash pattern=on 6pt off \offamount pt,
		line cap=round,
		line width = \standardlinewidth pt, 
		mark options={solid},
		line join=\linejoinchoice,
}}
\pgfplotsset{
	line4/.style={
		color4, 
		opacity=\standardlineopacity, 
		dash pattern=on 8pt off \offamount pt,
		line cap=round,
		line width = \standardlinewidth pt, 
		mark options={solid},
		line join=\linejoinchoice,
}}
\pgfplotsset{
	line5/.style={
		color5, 
		opacity=\standardlineopacity, 
		dash pattern=on 10pt off \offamount pt,
		line cap=round,
		line width = \standardlinewidth pt, 
		mark options={solid},
		line join=\linejoinchoice,
}}
\pgfplotsset{
	line6/.style={
		color7, 
		opacity=\standardlineopacity, 
		dash pattern=on 12pt off \offamount pt,
		line cap=round,
		line width = \standardlinewidth pt, 
		mark options={solid},
		line join=\linejoinchoice,
}}

\pgfplotsset{
	linegrad0/.style={
		color0!20!white, 
		opacity=\standardlineopacity, 
		line cap=round, 
		line width = \standardlinewidth + 2.5 pt, 
		mark options={solid},
		line join=\linejoinchoice,
}}

\pgfplotsset{
	linegrad1/.style={
		color0!45!white, 
		opacity=\standardlineopacity, 
		line cap=round, 
		line width = \standardlinewidth pt, 
		mark options={solid},
		line join=\linejoinchoice,
}}

\pgfplotsset{
	linegrad2/.style={
		color0!80!white, 
		opacity=\standardlineopacity, 
		line cap=round, 
		line width = \standardlinewidth pt, 
		mark options={solid},
		line join=\linejoinchoice,
}}

\pgfplotsset{
	linegrad3/.style={
		color0!85!black, 
		opacity=\standardlineopacity, 
		line cap=round, 
		line width = \standardlinewidth pt, 
		mark options={solid},
		line join=\linejoinchoice,
}}

\pgfplotsset{
	linegrad4/.style={
		color0!55!black, 
		opacity=\standardlineopacity, 
		line cap=round, 
		line width = \standardlinewidth pt, 
		mark options={solid},
		line join=\linejoinchoice,
}}

\pgfplotsset{
	linegrad5/.style={
		color0!10!black, 
		opacity=\standardlineopacity, 
		line cap=round, 
		line width = \standardlinewidth pt, 
		mark options={solid},
		line join=\linejoinchoice,
}}

\definecolor{color0}{rgb}{0.00392156862745098, 0.45098039215686275, 0.6980392156862745}
\definecolor{color1}{rgb}{0.8705882352941177, 0.5607843137254902, 0.0196078431372549}
\definecolor{color2}{rgb}{0.00784313725490196, 0.6196078431372549, 0.45098039215686275}
\definecolor{color3}{rgb}{0.8352941176470589, 0.3686274509803922, 0.0}
\definecolor{color4}{rgb}{0.8, 0.47058823529411764, 0.7372549019607844}
\definecolor{color5}{rgb}{0.792156862745098, 0.5686274509803921, 0.3803921568627451}
\definecolor{color6}{rgb}{0.984313725490196, 0.6862745098039216, 0.8941176470588236}
\definecolor{color7}{rgb}{0.5803921568627451, 0.5803921568627451, 0.5803921568627451}
\definecolor{color8}{rgb}{0.9254901960784314, 0.8823529411764706, 0.2}
\definecolor{color9}{rgb}{0.33725490196078434, 0.7058823529411765, 0.9137254901960784}

\colorlet{color0grad0}{color0!20!white}
\colorlet{color0grad1}{color0!45!white}
\colorlet{color0grad2}{color0!80!white}
\colorlet{color0grad3}{color0!85!black}
\colorlet{color0grad4}{color0!55!black}
\colorlet{color0grad5}{color0!10!black}

\colorlet{color3grad0}{color3!20!white}
\colorlet{color3grad1}{color3!45!white}
\colorlet{color3grad2}{color3!80!white}
\colorlet{color3grad3}{color3!85!black}
\colorlet{color3grad4}{color3!55!black}
\colorlet{color3grad5}{color3!10!black}

\usepackage{etoolbox}
\usepackage{siunitx}
\DeclareSIUnit\octave{oct}
\robustify\bfseries

\usepackage[acronym]{glossaries}
\newacronym{rosance}{ROSANCE}{RObust Signal And Noise Covariance Estimator}
\newacronym{ces}{CES}{complex elliptically symmetric}
\newacronym{scm}{SCM}{sample covariance matrix}
\newacronym{dof}{DoF}{degrees of freedom}

\newacronym{dft}{DFT}{discrete Fourier transform}
\newacronym{fft}{FFT}{fast Fourier transform}

\newacronym{rir}{RIR}{room impulse response}
\newacronym{iir}{IIR}{infinite impulse response}
\newacronym{fir}{FIR}{finite impulse response}
\newacronym{paf}{PAF}{plenacoustic function}

\newacronym{szc}{SZC}{sound zone control}
\newacronym{vast}{VAST}{variable span trade-off filter}
\newacronym{acc}{ACC}{acoustic contrast control}
\newacronym{pm}{PM}{pressure matching}

\newacronym{qos}{QoS}{quality of service}
\newacronym{fpi}{FPI}{fixed point iteration}

\newacronym{mse}{MSE}{mean square error}
\newacronym{nmse}{NMSE}{normalized mean square error}
\newacronym{ac}{AC}{acoustic contrast}
\newacronym{sd}{SD}{signal distortion}
\newacronym{re}{RE}{residual energy}
\newacronym{sinr}{SINR}{signal-to-interference-plus-noise ratio}
\newacronym{snr}{SNR}{signal-to-noise ratio}

\newacronym{evd}{EVD}{eigenvalue decomposition}
\newacronym{gevd}{GEVD}{generalized eigenvalue decomposition}
\newacronym{pevd}{PEVD}{polynomial eigenvalue decomposition}
\newacronym{pgevd}{PGEVD}{polynomial generalized eigenvalue decomposition}
\newacronym{svd}{SVD}{singular value decomposition}
\newacronym{mwf}{MWF}{multichannel Wiener filter}

\newacronym{ola}{OLA}{overlap-add}
\newacronym{ols}{OLS}{overlap-save}
\newacronym{wola}{WOLA}{weighted overlap-add}

\newacronym{lms}{LMS}{least mean squares}
\newacronym{nlms}{NLMS}{normalized least mean squares}
\newacronym{fxlms}{FxLMS}{filtered-x least mean squares}
\newacronym{rls}{RLS}{recursive least squares}

\newacronym{pseq}{PSEQ}{perfect sequence}
\newacronym{ops}{OPS}{orthogonal periodic sequence}
\newacronym{pps}{PPS}{perfect periodic sequence}

\newacronym{mnf}{MNF}{minimum noise fraction}
\newacronym{rkhs}{RKHS}{reproducing kernel Hilbert space}
\newacronym{krr}{KRR}{kernel ridge regression}

\newacronym{psd}{PSD}{positive semi-definite}
\newacronym{sdr}{SDR}{semi-definite relaxation}
\newacronym{sdp}{SDP}{semi-definite program}
\newacronym{qcqp}{QCQP}{quadratically constrained quadratic program}

\makeatletter
\newcommand*{\centerfloat}{%
	\parindent \z@
	\leftskip \z@ \@plus 1fil \@minus \textwidth
	\rightskip\leftskip
	\parfillskip \z@skip}
\makeatother

\usepackage{soul}

\DeclareMathOperator{\evalop}{\mathcal{M}}
\DeclareMathOperator{\linreg}{\mathcal{R}}
\newcommand{\rkhs}{\mathcal{H}}
\newcommand{\rkhstime}{\tilde{\mathcal{H}}}

\newcommand{\td}{\mathfrak{t}}
\newcommand{\fd}{\mathfrak{f}}

\newcommand{\fdmatrix}{\mathcal{B}(\fd)}
\newcommand{\tdmatrix}{\mathcal{B}(\td)}

\newcommand{\rdftlen}{L_f}

\newcommand{\pwcoeff}[1]{\mathring{#1}}

\begin{document}

	\title{Time-domain sound field estimation using kernel ridge regression}
	
	\author{Jesper Brunnström \orcidlink{0000-0003-2946-1268}, Martin Bo Møller \orcidlink{0000-0002-7323-7266}, Jan Østergaard \orcidlink{0000-0002-3724-6114}, Shoichi Koyama \orcidlink{0000-0003-2283-0884}, Toon van Waterschoot \orcidlink{0000-0002-6323-7350}, and Marc Moonen \orcidlink{0000-0003-4461-0073}
	\thanks{This research work was carried out at the ESAT Laboratory of KU Leuven, in the frame of Research Council KU Leuven C14-21-0075 "A holistic approach to the design of integrated and distributed digital signal processing algorithms for audio and speech communication devices", and has received funding from the European Union's Horizon 2020 research and innovation programme under the Marie Skłodowska-Curie grant agreement No. 956369: 'Service-Oriented Ubiquitous Network-Driven Sound — SOUNDS'. The scientific responsibility is assumed by the authors.}
	\thanks{Jesper Brunnström, Toon van Waterschoot, and Marc Moonen are with the Department of Electrical Engineering (ESAT), STADIUS, KU Leuven, Leuven, Belgium (e-mail: \url{jesper.brunnstroem@kuleuven.be})}
	\thanks{Martin Bo Møller is with Bang \& Olufsen, Acoustics R\&D, Struer, Denmark} 
	\thanks{Jan Østergaard is with the Department of Electronic Systems, Aalborg University, Aalborg, Denmark} 
	\thanks{Shoichi Koyama is with the National Institute of Informatics, Tokyo 101-8430, Japan}
	}

	\maketitle
	\begin{abstract}
		Sound field estimation methods based on kernel ridge regression have proven effective, allowing for strict enforcement of physical properties, in addition to the inclusion of prior knowledge such as directionality of the sound field. These methods have been formulated for single-frequency sound fields, restricting the types of data and prior knowledge that can be used. In this paper, the kernel ridge regression approach is generalized to consider discrete-time sound fields. The proposed method provides time-domain sound field estimates that can be computed in closed form, are guaranteed to be physically realizable, and for which time-domain properties of the sound fields can be exploited to improve estimation performance. Exploiting prior information on the time-domain behaviour of room impulse responses, the estimation performance of the proposed method is shown to be improved using a time-domain data weighting, demonstrating the usefulness of the proposed approach. It is further shown using both simulated and real data that the time-domain data weighting can be combined with a directional weighting, exploiting prior knowledge of both spatial and temporal properties of the room impulse responses. The theoretical framework of the proposed method enables solving a broader class of sound field estimation problems using kernel ridge regression where it would be required to consider the time-domain response rather than the frequency-domain response of each frequency separately.

		
		
	\end{abstract}


	\section{INTRODUCTION}
	\IEEEPARstart{S}{ound} field estimation is a fundamental audio signal processing problem, where the sound over a continuous spatial region is reconstructed from a limited set of measurements. Sound field estimation is crucially important for a number of sound field reproduction tasks, such as spatial active noise control, sound zone control, and spatial audio reproduction \cite{betlehemPersonal2015, zhangSurround2017}. Sound field estimation is difficult in part due to the cost of equipment and measurement time, generally leading to too few spatial samples being collected for perfect reconstruction \cite{ajdlerPlenacoustic2006}. As a consequence of the undersampling, sound field estimation is an ill-posed inverse problem, requiring prior knowledge to achieve satisfactory estimation performance. 
	

	A sound field produced by a spatially localized source in a room emitting an impulse is referred to as a \gls{rir} function, and such a function evaluated in a single point is referred to as a \gls{rir}. \Gls{rir} functions are of particular interest due to their importance in sound field reproduction applications. \Gls{rir} functions have a common structure that can be exploited \cite{vanwaterschootOptimally2008, bernschutzSound2012, florencioMaximum2015, jalmbyLowrank2021, helwaniGenerative2023}. Although such sound fields are mostly assumed to be stationary, the sound field is slowly time-varying \cite{elkoRoom2003, prawdaTime2023, olsenSound2017, prawdaShorttime2024}. The time-variation means that the quality of previous estimates decreases as time passes, leading to a need to re-measure, which in turn requires the sound field estimation process to be practical.

	

	A number of different approaches exist for sound field estimation, where different types of prior knowledge are exploited. If there exists a basis in which the sound field is sparse, the sound field can be reconstructed from fewer measurements, which is exploited in \cite{verburgReconstruction2018, katzbergCompressed2018, koyamaSparse2019, damianoCompressive2024}. A Bayesian approach allows for the inclusion of prior knowledge in the form of a probabilistic model \cite{caviedes-nozalSpatiotemporal2023, schmidSpatial2021}. Data-driven approaches have become more popular, which rely on prior knowledge extracted from data \cite{karakonstantisGenerative2023, miotelloDeep2024, shigemiPhysicsinformed2022, lluisSound2020, olivieriPhysicsinformed2024}. The sound field can also be modelled using a set of equivalent sources, with which prior knowledge in the form of sparsity  \cite{antonelloRoom2017} or an optimal mass transport regularization \cite{sundstromOptimal2024} can be exploited.
	
	
	The approach considered in this paper is based on \gls{krr} \cite{saburouTheory2016}. \Gls{krr} allows for optimization over a function space constrained to satisfy the Helmholtz equation, i.e. contains only physically realizable sound fields \cite{uenoKernel2018}. The \gls{krr} approach has been shown to be practical to apply to spatial active noise control \cite{itoFeedforward2019, brunnstromKernelinterpolationbased2021, koyamaSpatial2021}, sound zone control \cite{brunnstromVariable2022} and sound field reproduction \cite{koyamaWeighted2022}, in part due to the estimator being linear as a function of the data, which leads to practical algorithms when the estimators are integrated into the control methods. The \gls{krr} approach has been able to include more information about the sound field through a directional weighting \cite{uenoDirectionally2021, koyamaSpatial2021, brunnstromVariable2022, ribeiroSound2024}. The \gls{krr} approach has also been shown to be closely related to Bayesian inference of spherical harmonic wave functions \cite{samarasingheWavefield2014, samarasingheEfficient2015, uenoSound2018, brunnstromBayesian2024, brunnstromBayesian2025}, as well as Gaussian process regression \cite{rasmussenGaussian2006, caviedes-nozalGaussian2021, fernandez-grandeReconstruction2021, fengRoom2024, liangSound2024}. The strength of the approach lies in the representer theorems \cite{scholkopfGeneralized2001, dinuzzoRepresenter2012, argyriouUnifying2014, minhUnifying2016, diwaleGeneralized2018, boyerRepresenter2019}, which characterize the optimal solutions of a large class of \gls{krr} problems, reducing infinite-dimensional optimization problems to finite-dimensional problems. 
	
	
	



	Existing sound field estimation methods based on \gls{krr} \cite{uenoSound2018, uenoDirectionally2021, ribeiroSound2024} consider single-frequency sound fields. This restricts the set of problems that can be solved to those which can be formulated for each frequency individually, in addition to limiting the type of prior knowledge that can be used. Although a continuous-time formulation was proposed in \cite{sundstromSound2024}, in contrast to other kernel-based methods, no closed form solution is available, making the method costly to use. In addition, the obtained solution only satisfies the wave equation approximately by minimizing a physics-informed cost, rather than guaranteeing that such a constraint is exactly satisfied. 
	
	A generalization of the \gls{krr} approach is proposed in this paper, with which discrete-time sound fields in continuous space are modelled. The proposed method provides estimates which are guaranteed to satisfy the wave equation with a kernel that is obtained in closed form. Using the \gls{dft}, a \gls{rkhs} modelling discrete-time sound fields is derived. The model allows for the inclusion of prior knowledge both relating to the time-domain and frequency-domain properties of the sound field. This is shown to be useful for the identification of \glspl{rir}, which have a high degree of predictable time-domain behaviour. 
	
	The paper makes a theoretical contribution, by constructing sound field models and function spaces with which sound field-related optimization problems beyond those considered in the paper can be solved with \gls{krr}. Compared to previous methods, the generalization to time-domain sound fields allows for a broader class of problems to be considered using \gls{krr}. Such problems include those where the sound field at multiple frequencies must be considered jointly, or where the time-domain response is of crucial importance. 

	The structure of the paper is as follows. In section~\ref{sec:problem-statement} the sound field estimation problem is introduced, with a model of the available data. In section~\ref{sec:sound-field-model} the sound field model is introduced, and a \gls{rkhs} modelling time-domain sound fields is defined. In section~\ref{sec:sound-field-estimation} the kernel ridge regression problem is defined and solved, producing a closed form expression for the optimal sound field estimate. In section~\ref{sec:regularization-and-data-weighting} the inclusion of prior knowledge is considered, introducing a directionally weighted regularization, and a data-weighting using the time-domain properties of the \glspl{rir}. In section~\ref{sec:evaluation} the proposed method is evaluated in a number of experiments. Finally the paper is concluded in section~\ref{sec:conclusion}. To facilitate easy use of the results, code associated with the proposed method is made available at \href{https://github.com/sounds-research/aspcol}{github.com/sounds-research/aspcol}.


	\subsection{Notation}
	The vector space of $N$-dimensional real vectors is denoted by $\real^{N}$, the vector space of $N$-dimensional complex vectors is denoted by $\complex^{N}$. The unit sphere $\{\bm{x} \;\vert \;\lVert \bm{x} \rVert = 1, \bm{x} \in \real^3\}$ in three dimensions, i.e. the surface of the unit ball, is denoted by $\unitsphere$. Finite-dimensional vectors are denoted by lower-case bold letters such as $\bm{a}$, matrices by upper-case bold letters such as $\bm{A}$, and scalar values by non-bold letters such as $a$. The complex conjugate of a variable $a$ is denoted by $\conj{a}$. The transpose of a matrix $\bm{A}$ is denoted by $\bm{A}^\top$, and the Hermitian transpose by $\bm{A}^\herm$. The imaginary unit is written as $\imag^2 = -1$. A linear operator operates on the variable to the right of it, e.g. the variable $a$ is the argument of the linear operator $T$ in the expression $T a$. For a linear operator $T : \mathcal{H} \rightarrow \mathcal{H}'$ between Hilbert spaces $\mathcal{H}$ and $\mathcal{H}'$, its unique adjoint is denoted by $T^{*} : \mathcal{H}' \rightarrow \mathcal{H}$. Selecting an element with index $l$ from $\bm{a}$ is written as $(\bm{a})_l$.
	
	
	
	\section{Problem statement}\label{sec:problem-statement}
	\subsection{Data model}
	The objective is to estimate the sound field from the available data, as will be described in this section. Consider a simply connected region $\Omega \subset \real^3$ in a room within which the \glspl{rir} from one or more sources should be estimated. Within the region $\Omega$ there are $M$ omnidirectional microphones placed at the positions $\bm{r}_1, \bm{r}_2, \dots, \bm{r}_M \in \Omega$. The \gls{rir} associated with a single source and a single position $\bm{r}$ can be modelled as an $L$-length sequence, representing a \gls{fir} filter. The set of all such sequences is denoted by $\mathfrak{t} = \real^L$, which is considered an inner product space with the standard Euclidean inner product $\langle \tilde{\bm{a}},  \tilde{\bm{b}} \rangle_{\td} = \sum_{n=0}^{L-1} (\tilde{\bm{b}})_n (\tilde{\bm{a}})_n = \tilde{\bm{b}}^\top \tilde{\bm{a}}$. 
	
	The \gls{rir} associated with a single source for all positions $\bm{r} \in \Omega$ can then be represented by a vector-valued function $\tilde{\bm{u}} : \Omega \rightarrow \td$, which corresponds to the sound field that should be estimated. The signal recorded by a microphone placed at position $\bm{r}_m \in \Omega$ at discrete-time index $n \in \integer$ is
	\begin{equation}
		p(\bm{r}_m, n) = \langle \tilde{\bm{u}}(\bm{r}_m), \tilde{\bm{\phi}}(n) \rangle_{\mathfrak{t}} + s_m(n),
		\label{eq:data-model-sound-pressure}
	\end{equation}
	where $\tilde{\bm{\phi}}(n) \in \mathfrak{t}$ denotes a vector containing the $L$ most recent values of the source signal, and $s_m$ represents the noise. It is assumed that there is one source of interest active at any given time for which the source signal $\phi$ is known, and any other sound in the recording is considered part of the noise $s_m$. The assumption holds for instance when estimating the \gls{rir} of a loudspeaker under the user's control. 
	
	If the signal in \eqref{eq:data-model-sound-pressure} is collected for a sufficient number of samples for each microphone, the measured signal can be deconvolved to obtain
	\begin{equation}
		\tilde{\bm{h}}_m = \tilde{\bm{u}}(\bm{r}_m) + \tilde{\bm{\epsilon}}_m,
		\label{eq:data-model}
	\end{equation}
	 where $\tilde{\bm{\epsilon}}_m \in \td$ is an additive error that can depend on the measurement noise $s$, source signal $\phi$, and the deconvolution process. The available data associated with microphone $m$ is therefore the vector $\tilde{\bm{h}}_m \in \td$. The task considered in this paper is to estimate the \gls{rir} function $\tilde{\bm{u}}$ from data of the form \eqref{eq:data-model}. 
	


	\subsection{Frequency-domain representation} \label{sec:frequency-domain-representation}
	

	To relate the time-domain signals to their frequency-domain characteristics, the \gls{dft} is used. The conventional \gls{dft} contains redundancy when applied to real-valued signals, expressed as a conjugate symmetry in the frequency-domain signals. The redundancy can be eliminated by only considering the first $\rdftlen = \lfloor \frac{L}{2}\rfloor + 1$ frequency-domain values. The set of frequency-domain signals is then defined as $\fd = \complex^{\rdftlen - 2} \times \real^2$ if $L$ is even and $\fd = \complex^{\rdftlen-1} \times \real$ if $L$ is odd, due to the frequencies at $l=0$ and $l=\frac{L}{2}$ being strictly real-valued, where $l$ is the discrete frequency index. The \gls{dft} $\mathcal{F} : \td \rightarrow \fd$ is defined as\footnote{The time convention is chosen to be consistent with the acoustics literature \cite{martinMultiple2006}, but is opposite to some popular implementations \cite{bradburyJAX2018, harrisArray2020}.}
	\begin{equation}
		(\mathcal{F} \tilde{\bm{a}})_l = \sum_{n=0}^{L-1} e^{2 \pi \imag n l  / L} (\tilde{\bm{a}})_n \quad 0 \leq l < \rdftlen,
	\end{equation}
	with the inverse \gls{dft} $\mathcal{F}^{-1} : \fd \rightarrow \td$ defined as 
	\begin{equation}
		(\mathcal{F}^{-1} \bm{a})_n = \mathfrak{Re} \biggl[\sum_{l = 0}^{\rdftlen-1} c_l e^{-2 \pi \imag n l / L} (\bm{a})_l \biggr] \quad 0 \leq n < L,
	\end{equation}
	where the real-part operator is defined as $\mathfrak{Re} [a] = \frac{1}{2} (a + \conj{a})$ for any vector $a$, and
	\begin{equation}
		c_l = \begin{cases}
			\frac{2}{L} & \quad \text{if } 0 < l < \frac{L}{2} \\
			\frac{1}{L} & \quad \text{if } l=0 \text{ or } l=\frac{L}{2}.
		\end{cases}
	\end{equation}
	The forward transform can be written in matrix form as $\bm{a} = \bm{F} \tilde{\bm{a}}$ where $\bm{F} \in \complex^{\rdftlen \times L}$ defined by $(\bm{F})_{ln} = e^{2 \pi \imag n l / L}$. The inverse transform is $\tilde{\bm{a}} = \mathfrak{Re}[\bm{B} \bm{C} \bm{a}]$, where $\bm{B} \in \complex^{L \times \rdftlen}$ is defined by $(\bm{B})_{nl} = e^{-2 \pi \imag n l / L}$ and the diagonal matrix $\bm{C} \in \real^{L_f \times L_f}$ is defined by $(\bm{C})_{ll} = c_l$.

	It is desirable for the transform to be unitary, like the conventional \gls{dft} is up to a normalization constant when considered on spaces with the standard Euclidean inner product. This property leads to equivalent inner products and norms of the time and frequency-domain representations, specifically $\langle \tilde{\bm{a}}, \tilde{\bm{b}} \rangle_{\td} = \langle \mathcal{F} \tilde{\bm{a}}, \mathcal{F} \tilde{\bm{b}} \rangle_{\fd}$. A unitary transform satisfies $\mathcal{F}^{*} = \mathcal{F}^{-1}$, where $\mathcal{F}^{*} : \fd \rightarrow \td$ is the Hilbert space adjoint defined by $\langle\mathcal{F} \tilde{\bm{a}}, \bm{b} \rangle_{\fd} = \langle\tilde{\bm{a}}, \mathcal{F}^{*}\bm{b} \rangle_{\td}$ for all $\tilde{\bm{a}} \in \td, \bm{b} \in \fd$. The unitary property of $\mathcal{F}$ can be used to determine the inner product of $\mathfrak{f}$ as
	\begin{equation}
		\langle \bm{a},  \bm{b} \rangle_{\fd} = \sum_{l = 0}^{\rdftlen - 1} c_l \mathfrak{Re}[(\bm{a})_l \conj{(\bm{b})}_l] =  \mathfrak{Re}[\bm{b}^{\herm} \bm{C} \bm{a}].
		\label{eq:freq-domain-inner-product}
	\end{equation}
	The transform $\mathcal{F}$ is then linear, bijective, and unitary. Linear operators on $\td$ and $\fd$, the sets of which are referred to as $\tdmatrix$ and $\fdmatrix$ respectively, are characterized in Appendix \ref{sec:discrete-fourier-transform}. 
	
	

	
	\section{Sound field model}\label{sec:sound-field-model}
	In this section, the goal is to construct a function space containing physically realizable sound fields $\tilde{\bm{u}}$. The \gls{rkhs} of \cite{uenoKernel2018, uenoDirectionally2021} which is defined for a single frequency will be adapted to the \gls{dft}-based frequency-domain representation in Section~\ref{sec:frequency-domain-representation} to define a \gls{rkhs} representing time-domain sound fields in Section~\ref{sec:definition-time-domain-rkhs}. 
	
	
\subsection{Frequency-domain plane wave model}
	Consider a sound field in a source-free region $\Omega \subset \real^3$, represented by the sound pressure $u_{\omega} : \Omega \rightarrow \complex$, a function of the position $\bm{r} \in \Omega$ for a specific angular frequency $\omega \in \real_{\geq 0}$. Such a sound field in a source-free region satisfies the homogenous Helmholtz equation 
	\begin{equation}
		\Bigl(\Delta + \frac{\omega^2}{c^2} \Bigr) u_\omega = 0,
		\label{eq:helmholtz-equation}
	\end{equation}
	where $\Delta$ is the Laplace operator, and $c \in \real_{> 0}$ is the speed of sound, assumed to be constant in $\Omega$. 
	
	
	A solution to \eqref{eq:helmholtz-equation} can be approximated arbitrarily well with the Herglotz integral \cite[Appendix A]{uenoDirectionally2021}, defined as 
	\begin{equation}
		u_\omega(\bm{r}) = \int_{\unitsphere} e^{-\imag \frac{\omega}{c} \bm{r}^\top \unitvec{\bm{d}}}  \pwcoeff{u}_\omega(\unitvec{\bm{d}})\,ds(\unitvec{\bm{d}}).
		\label{eq:herglotz}
	\end{equation}
	The directionality function $\pwcoeff{u}_\omega : \unitsphere \rightarrow \complex$ is a square integrable function, the space of which is denoted by $L^2(\unitsphere \rightarrow \complex)$, within which functions are considered equivalent if they only differ on a set of measure zero \cite[Section 2.4.6]{kennedyHilbert2013}. The directionality function evaluated for a single direction $\unitvec{\bm{d}}$ can be interpreted as the complex coefficient of a plane wave \cite{williamsFourier1999, uenoDirectionally2021, ribeiroSound2024, brunnstromBayesian2025}. The integral is formulated in terms of the natural spherical measure $s$, and $\unitvec{\bm{d}} \in \unitsphere$ corresponds to the incoming plane wave direction, which is the opposite of the propagation direction.

	\subsection{Time-domain plane wave model}
	The sound field model in \eqref{eq:herglotz} for continuous frequencies $\omega \in \real$ should be adapted to the function $\bm{u} : \Omega \rightarrow \fd$ representing a sound field for the $\rdftlen$ sampled frequencies. The proposed sound field model is
	\begin{equation}
		\bm{u}(\bm{r}) = \int_{\unitsphere} \bm{E}(\bm{r}, \unitvec{\bm{d}}) \pwcoeff{\bm{u}}(\unitvec{\bm{d}}) \,ds(\unitvec{\bm{d}}),
		\label{eq:frequency-domain-plane-wave-decomposition}
	\end{equation}
	where $\bm{E}(\bm{r}, \unitvec{\bm{d}}) \in \fdmatrix$ contain the plane wave basis functions, and the directionality function $\pwcoeff{\bm{u}} : \unitsphere \rightarrow \fd$ describes the plane wave coefficients at all sampled frequencies. The basis functions $\bm{E}(\bm{r}, \unitvec{\bm{d}})$ can be represented by the diagonal matrix 
	\begin{equation}
		(\bm{E}(\bm{r}, \unitvec{\bm{d}}))_{ll} = \begin{cases}
			e^{-\imag \frac{\omega_l}{c}\bm{r}^\top \unitvec{\bm{d}} } & 0 \leq l < \frac{L}{2} \\
			\cos(\frac{\omega_l}{c}\bm{r}^\top \unitvec{\bm{d}}) & \text{if } l = \frac{L}{2}.
		\end{cases}
	\label{eq:plane-wave-sampled}
	\end{equation}
	The sampled angular frequency $\omega_l$ corresponds to the continuous angular frequency $\omega_l = \qty[parse-numbers=false]{\frac{2 \pi f_s l}{L}}{\radian / \second}$ where $f_s$ is the sampling rate in \unit{\hertz}.

	Due to the real-value constraint at $l=\frac{L}{2}$, the standard plane wave is exchanged for $\mathfrak{Re}[e^{-\imag \frac{\omega_l}{c}\bm{r}^\top \unitvec{\bm{d}}}] = \cos(\frac{\omega_l}{c}\bm{r}^\top \unitvec{\bm{d}})$. The real-value constraint is satisfied without changes for $l=0$. This means that the physical correctness of the model is lost for $l = \frac{L}{2}$, in order to maintain the properties of $\fd$ as sequences obtained from the \gls{dft}. However, this is not considered to be a significant drawback in practice. Whenever $L$ is odd, there is no frequency with index $\frac{L}{2}$. Otherwise if $L$ is even, given that any measurement is processed with an ideal anti-aliasing filter having a zero output at the Nyquist frequency, no error is incurred.

	
	A time-domain sound field that is either $L$ samples in length, or periodic with period $L$ can be represented by a function $\tilde{\bm{u}} : \Omega \rightarrow \td$. Any such sound field can be defined via the \gls{dft} as 
	\begin{equation}
		\tilde{\bm{u}}(\bm{r}) = \mathcal{F}^{-1} \bm{u}(\bm{r}).
		\label{eq:time-domain-sound-field-function}
	\end{equation}
	Due to the invertibility of the \gls{dft}, both a frequency-domain and time-domain sound field can be represented by the same directionality function $\pwcoeff{\bm{u}} : \unitsphere \rightarrow \fd$. 
	



	\subsection{Reproducing kernel Hilbert space for frequency-domain sound fields}
	Following \cite{uenoKernel2018, uenoDirectionally2021}, the function space for sound fields $u_\omega : \Omega \rightarrow \complex$ of a single frequency $\omega$ can be represented by
	\begin{equation}
		\mathcal{H}_{\omega} = \{\mathcal{A}_{\omega} \pwcoeff{u}_\omega \vert \pwcoeff{u}_\omega \in L^2(\unitsphere \rightarrow \complex)\},
		\label{eq:single-freq-rkhs}
	\end{equation}
	where $\mathcal{A}_\omega$ is the operator transforming a function $\pwcoeff{u}_\omega$ to $u_\omega$ as in \eqref{eq:herglotz}. The associated inner product is 
		\begin{equation}
		\begin{aligned}
			\langle u_\omega, v_\omega \rangle_{\rkhs_{\omega}} &= \int_{\unitsphere} \pwcoeff{u}_\omega(\unitvec{\bm{d}}) \conj{\pwcoeff{v}_\omega(\unitvec{\bm{d}})} \,ds(\unitvec{\bm{d}}).
			\label{eq:single-frequency-inner-product}
		\end{aligned}
	\end{equation}
	The Hilbert space $\mathcal{H}_{\omega}$ is a \gls{rkhs}, and is characterized by the kernel function $\kappa_{\omega} : \Omega \times \Omega \rightarrow \complex$ defined as $\kappa_{\omega}(\bm{r}, \bm{r}') = j_0\bigl(\frac{\omega}{c} \lVert \bm{r} - \bm{r}' \rVert_2 \bigr)$, where $j_0$ is the zeroth order spherical Bessel function of the first kind \cite{uenoKernel2018}.
	
%

	An analogous function space for sound fields  $\bm{u} : \Omega  \rightarrow \fd$ on the considered set of sampled frequencies can be defined as
		\begin{equation}
		\mathcal{H} = \{\mathcal{A} \pwcoeff{\bm{u}} \vert \pwcoeff{\bm{u}} \in L^2(\unitsphere \rightarrow \fd)\},
		\label{eq:multi-freq-rkhs}
	\end{equation}
	where $\mathcal{A}$ is the operator transforming functions $\pwcoeff{\bm{u}}$ to $\bm{u}$ corresponding to \eqref{eq:frequency-domain-plane-wave-decomposition}. The associated inner product analogous to \eqref{eq:single-frequency-inner-product} can be defined as 
\begin{equation}
	\begin{aligned}
		\langle \bm{u}, \bm{v} \rangle_{\rkhs} = \int_{\unitsphere} \langle \pwcoeff{\bm{u}}(\unitvec{\bm{d}}), \pwcoeff{\bm{v}}(\unitvec{\bm{d}}) \rangle_{\fd} \,ds(\unitvec{\bm{d}}).
		\label{eq:multi-freq-rkhs-inner-product}
	\end{aligned}
\end{equation}
The kernel function of the \gls{rkhs} $\rkhs$ is a positive definite function $\Gamma : \Omega \times \Omega \rightarrow \fdmatrix$, which can be represented by the diagonal matrix
\begin{equation}
		\begin{aligned}
	(\Gamma(\bm{r}, \bm{r}'))_{ll} &= 
	\begin{cases}
		\kappa_{\omega_l}(\bm{r}, \bm{r}') & 0 \leq l < \frac{L}{2} \\
		\frac{1}{2} (\kappa_{\omega_l}(\bm{r}, \bm{r}') + \kappa_{\omega_l}(\bm{r}, -\bm{r}') ) & l = \frac{L}{2},
	\end{cases}
\end{aligned}
	\label{eq:multi-freq-rkhs-kernel}
\end{equation}
as is shown in appendix \ref{sec:frequency-domain-rkhs-details}.


%

	\subsection{Reproducing kernel Hilbert space for time-domain sound fields}\label{sec:definition-time-domain-rkhs}
	There is a unique time-domain sound field $\tilde{\bm{u}} \in \rkhstime$ for each frequency-domain sound field $\bm{u} \in \mathcal{H}$, due to the invertibility of the \gls{dft}. The set of time-domain sound fields can therefore be defined in terms of the frequency-domain sound fields as 
	\begin{equation}
		\tilde{\mathcal{H}} = \{ \tilde{\bm{u}} \;\vert \; \tilde{\bm{u}}(\bm{r}) = \mathcal{F}^{-1} \bm{u}(\bm{r}), \;\forall \bm{r} \in \Omega,  \bm{u} \in \mathcal{H}\}.
	\end{equation}
	Due to the equivalence of the time- and frequency-domain representations, the natural choice of inner product is the same as for $\rkhs$ in \eqref{eq:multi-freq-rkhs-inner-product}, specifically, $\langle \tilde{\bm{u}}, \tilde{\bm{v}} \rangle_{\rkhstime} = \langle \bm{u}, \bm{v} \rangle_{\rkhs}$. The reproducing kernel $\tilde{\Gamma} : \Omega \times \Omega \rightarrow \tdmatrix$ is obtained in terms of \eqref{eq:multi-freq-rkhs-kernel} as 
	\begin{equation}
		\tilde{\Gamma}(\bm{r}, \bm{r}') = \mathcal{F}^{-1} \Gamma(\bm{r}, \bm{r}') \mathcal{F} = \mathfrak{Re}[\bm{B} \bm{C} \Gamma(\bm{r}, \bm{r}') \bm{F}],
		\label{eq:time-domain-kernel-function}
	\end{equation}
	as is shown in appendix \ref{sec:time-domain-rkhs-details}. 
	
	
	
	\section{Sound field estimation}\label{sec:sound-field-estimation}
	\subsection{Kernel ridge regression}
	To estimate a sound field, the task is to find a function in $\rkhstime$ that corresponds to data in the form of \eqref{eq:data-model}. Since $\rkhstime$ is a \gls{rkhs}, the task can be accomplished through \gls{krr}. An optimization problem can then be formulated as
	\begin{equation}
		\min_{\tilde{\bm{u}} \in \rkhstime} \sum_{m=1}^{M}\lVert \tilde{\bm{h}}_m - \tilde{\bm{u}}(\bm{r}_m) \rVert_{\td}^2 + \lambda \lVert \tilde{\bm{u}} \rVert_{\rkhstime}^2,
		\label{eq:krr-stationary-microphones}
	\end{equation}
	where $\lambda \in \real_{\geq 0}$ is a regularization parameter. 
	
	Due to the representer theorem \cite{diwaleGeneralized2018}, the optimal solution of \eqref{eq:krr-stationary-microphones} evaluated at any $\bm{r} \in \Omega$ has the form
	\begin{equation}
		\tilde{\bm{u}}^{\text{opt}} (\bm{r}) = \sum_{m=1}^{M} \tilde{\Gamma}(\bm{r}, \bm{r}_m) \tilde{\bm{a}}_m
		\label{eq:optimal-form-stationary-microphones}
	\end{equation}
	for a set of unknown parameters $\tilde{\bm{a}}_m \in \td$ for $m = 1, 2, \dots, M$. Inserting \eqref{eq:optimal-form-stationary-microphones} into \eqref{eq:krr-stationary-microphones}, an equivalent finite-dimensional optimization problem is obtained as 
	\begin{equation}
		\min_{\tilde{\bm{a}} \in \td^M} \; \langle (\tilde{\Gamma}^2 + \lambda \tilde{\Gamma}) \tilde{\bm{a}}, \tilde{\bm{a}} \rangle_{\td^M} - 2 \langle \tilde{\Gamma} \tilde{\bm{a}}, \tilde{\bm{h}} \rangle_{\td^M} + \langle\tilde{\bm{h}}, \tilde{\bm{h}} \rangle_{\td^M},
		\label{eq:finite-dim-optimization-problem}
	\end{equation}
	where the space $\td^M$ is the direct sum $\bigoplus_{m=1}^{M} \td$ with inner product $\langle \tilde{\bm{a}}, \tilde{\bm{b}} \rangle_{\td^M} = \sum_{m=1}^{M} \langle \tilde{\bm{a}}_m, \tilde{\bm{b}}_m \rangle_{\td}$, corresponding to $\real^{ML}$. The stacked vectors in \eqref{eq:finite-dim-optimization-problem} are
	\begin{equation}
		\begin{aligned}
			\tilde{\bm{a}} &= (\tilde{\bm{a}}_1, \dots, \tilde{\bm{a}}_M) \in \td^M \\
			 \tilde{\bm{h}} &= (\tilde{\bm{h}}_1, \dots, \tilde{\bm{h}}_M) \in \td^M ,
		\end{aligned}
	\end{equation}
	and the stacked kernel $\tilde{\Gamma} \in \mathcal{B}(\td^M)$ is represented by the matrix
		\begin{equation}
		\tilde{\Gamma} = \begin{bmatrix} 
			\tilde{\Gamma}(\bm{r}_1, \bm{r}_1) & \dots & \tilde{\Gamma}(\bm{r}_1, \bm{r}_M) \\
			\vdots& \ddots & \vdots\\
			\tilde{\Gamma}(\bm{r}_M, \bm{r}_1) & \dots & \tilde{\Gamma}(\bm{r}_M, \bm{r}_M) \\
		\end{bmatrix}.
	\end{equation}
	Note that $\tilde{\Gamma}$ is self-adjoint, due to $\tilde{\Gamma}(\bm{r}, \bm{r}') = \tilde{\Gamma}(\bm{r}', \bm{r})^{*}$.


	By computing the gradient of \eqref{eq:finite-dim-optimization-problem} and setting it to zero, and by observing that the problem is convex, the unique optimum is obtained, which is
	\begin{equation}
		\begin{aligned}
			\tilde{\bm{a}}^{\text{opt}} &= (\tilde{\Gamma} + \lambda \bm{I})^{-1} \tilde{\bm{h}}.
			\label{eq:optimum-krr-stationary-microphones}
		\end{aligned}
	\end{equation}
	The optimal parameters \eqref{eq:optimum-krr-stationary-microphones} can be inserted in \eqref{eq:optimal-form-stationary-microphones} to obtain an estimate of the sound field for any position $\bm{r} \in \Omega$. The solution in \eqref{eq:optimum-krr-stationary-microphones} is equivalent to the frequency-domain solution in \cite{uenoKernel2018}, as is shown in Appendix~\ref{sec:time-frequency-equivalence}. 
	
	
	\subsection{Kernel ridge regression with prior knowledge}
	Obtaining a sufficiently large number of spatial samples is difficult in practice, hence it is important to incorporate prior knowledge into the optimization problem, in order to maintain estimation performance despite undersampling. This can be done in the \gls{krr} framework by adding an appropriate regularization. 
	
	
	A more general \gls{krr} formulation than \eqref{eq:krr-stationary-microphones} is proposed here, defined as
	\begin{equation}
		\min_{\tilde{\bm{u}} \in \rkhstime} \sum_{m=1}^{M}\lVert \tilde{\bm{h}}_m- \tilde{\bm{u}}(\bm{r}_m) \rVert_{\tilde{\bm{Q}}_{m}}^2 + \lambda \lVert \linreg \tilde{\bm{u}} \rVert_{\mathcal{Z}}^2,
		\label{eq:krr-stationary-microphones-regularized}
	\end{equation}
	where $\linreg : \rkhstime \rightarrow \mathcal{Z}$ is a linear operator, $\mathcal{Z}$ is an arbitrary Hilbert space, and $\lVert \cdot \rVert_{\tilde{\bm{Q}}_{m}}$ is a weighted norm induced by the inner product $\langle \tilde{\bm{a}}, \tilde{\bm{b}} \rangle_{\tilde{\bm{Q}}_{m}} = \langle \tilde{\bm{Q}}_{m} \tilde{\bm{a}}, \tilde{\bm{b}} \rangle_{\td}$ for a self-adjoint positive definite linear operator $\tilde{\bm{Q}}_m \in \tdmatrix$. The linear operator $\linreg$ should penalize functions that are unlikely to represent the true sound field by increasing their norm. Functions that closely match the prior knowledge should be given a small norm, promoting them as optimal solutions. The data weighting defined by $\tilde{\bm{Q}}_m$ allows for e.g. different levels of importance to be given to different parts of the data. How to choose $\tilde{\bm{Q}}_m$ and $\linreg$ will be investigated in Section~\ref{sec:regularization-and-data-weighting}. 
	
	In order to solve \eqref{eq:krr-stationary-microphones-regularized} in a straightforward manner, the regularization operator $\linreg$ is restricted to be invertible with domain $\rkhstime$. The regularization term can then be written as $\lVert \linreg \tilde{\bm{u}} \rVert_{\mathcal{Z}}^2 = \langle \linreg \tilde{\bm{u}}, \linreg \tilde{\bm{u}} \rangle_{\mathcal{Z}} = \langle \tilde{\bm{u}}, \tilde{\bm{u}} \rangle_{\rkhstime_{\linreg}} = \lVert \tilde{\bm{u}} \rVert_{\rkhstime_{\linreg}}^2$, where $\rkhstime_{\linreg}$ is an inner product space over the same set of functions as $\rkhstime$, but with the inner product $\langle \tilde{\bm{u}}, \tilde{\bm{v}} \rangle_{\rkhstime_{\linreg}} =  \langle \linreg \tilde{\bm{u}}, \linreg \tilde{\bm{v}} \rangle_{\mathcal{Z}}$. A necessary implication of this assumption on $\linreg$ is that $\mathcal{Z}$ has to be infinite-dimensional. The optimization problem in \eqref{eq:krr-stationary-microphones-regularized} is then equivalent to
	\begin{equation}
		\min_{\tilde{\bm{u}} \in \rkhstime_{\linreg}} \sum_{m=1}^{M}\lVert \tilde{\bm{h}}_m - \tilde{\bm{u}}(\bm{r}_m) \rVert_{\tilde{\bm{Q}}_{m}}^2 + \lambda \lVert \tilde{\bm{u}} \rVert_{\rkhstime_{\linreg}}^2,
		\label{eq:krr-stationary-microphones-regularized-HR}
	\end{equation}

	The kernel function $\tilde{\Gamma}_{r} : \Omega \times \Omega \rightarrow \tdmatrix$ associated with $\rkhstime_{\linreg}$ is defined by
	\begin{equation}
		\tilde{\Gamma}_{r}(\bm{r}, \bm{r}') \tilde{\bm{c}} = \Bigl( (\linreg^{*} \linreg)^{-1} \tilde{\Gamma}(\cdot, \bm{r}') \tilde{\bm{c}} \Bigl)(\bm{r})
		\label{eq:kernel-function-HR}
	\end{equation}
	for all $\tilde{\bm{c}} \in \td$, as is shown in appendix \ref{sec:regularized-kernel-function}.

	Due to the representer theorem \cite{diwaleGeneralized2018}, the optimal solution of \eqref{eq:krr-stationary-microphones-regularized-HR} evaluated at an arbitrary $\bm{r} \in \Omega$ can be written in the form 
	\begin{equation}
		\tilde{\bm{u}}^{\text{opt}}(\bm{r}) = \sum_{m=1}^{M} \tilde{\Gamma}_{r}(\bm{r}, \bm{r}_m) \tilde{\bm{a}}_m
		\label{eq:optimal-form-stationary-microphones-regularized-operator}
	\end{equation}
	for the unknown parameters $\tilde{\bm{a}}_m \in \td$ for $m = 1, \dots, M$.

	Inserting \eqref{eq:optimal-form-stationary-microphones-regularized-operator} into \eqref{eq:krr-stationary-microphones-regularized-HR} an equivalent finite-dimensional optimization problem is obtained as
		\begin{equation}
		\begin{aligned}
			\min_{\tilde{\bm{a}} \in \td^M} &= \langle (\tilde{\Gamma}_r \tilde{\bm{Q}} \tilde{\Gamma}_r + \lambda \tilde{\Gamma}_{r})\tilde{\bm{a}}, \tilde{\bm{a}}  \rangle_{\td^M} \\
			&- 2\langle \tilde{\bm{Q}} \tilde{\Gamma}_r \tilde{\bm{h}}, \tilde{\bm{a}} \rangle_{\td^M} + \langle \tilde{\bm{Q}} \tilde{\bm{h}}, \tilde{\bm{h}} \rangle_{\td^M},
		\end{aligned}
	\label{eq:finite-dim-optimization-problem-regularized}
	\end{equation}
	where $\tilde{\bm{Q}} = \diag\{\tilde{\bm{Q}}_m\}_{m=1}^{M} \in \mathcal{B}(\td^M)$, with $\diag$ defined such that $\tilde{\bm{Q}} \tilde{\bm{a}} = (\tilde{\bm{Q}}_1 \tilde{\bm{a}}_1, \dots, \tilde{\bm{Q}}_M \tilde{\bm{a}}_M)$. Both $\tilde{\bm{Q}}$ and $\tilde{\Gamma}_r$ are self-adjoint, which was used to simplify the expression. 
	


	By computing the gradient of \eqref{eq:finite-dim-optimization-problem-regularized} and setting it to zero, the unique optimum is obtained, which is
	\begin{equation}
		\begin{aligned}
			\tilde{\bm{a}}^{\text{opt}} &= (\tilde{\Gamma}_r + \lambda \tilde{\bm{Q}}^{-1} )^{-1} \tilde{\bm{h}},
		\end{aligned}
	\label{eq:optimal-a-regularized}
	\end{equation}
	and can be inserted into \eqref{eq:optimal-form-stationary-microphones-regularized-operator} to obtain the sound field estimate for any $\bm{r} \in \Omega$.

	\section{Regularization and data weighting}\label{sec:regularization-and-data-weighting}
	\subsection{Directional regularization}\label{sec:directional-regularization}
	An informative choice of regularization operator $\linreg$ can improve the estimate in \eqref{eq:optimal-a-regularized}. It has been shown in \cite{uenoDirectionally2021, koyamaSpatial2021, brunnstromVariable2022, ribeiroSound2024} that a directional weighting can significantly improve estimation performance by exploiting prior knowledge of the propagation direction of the sound field. An analogous but generalized directional weighting can be used in the proposed \gls{rkhs} $\rkhstime$. The directional weighting operator $\mathcal{R} : \rkhstime \rightarrow \rkhstime$ can be defined as
	\begin{equation}
		(\linreg \tilde{\bm{u}})(\bm{r}) = \mathcal{F}^{-1} \int_{\unitsphere} \bm{E}(\bm{r}, \unitvec{\bm{d}}) \bm{W}(\hat{\bm{d}}) \pwcoeff{\bm{u}}(\hat{\bm{d}}) ds(\hat{\bm{d}}),
		\label{eq:directional-weighting-operator}
	\end{equation}
	where $\bm{W}(\unitvec{\bm{d}}) : \unitsphere \rightarrow \fdmatrix$. The adjoint operator $\linreg^{*} : \rkhstime \rightarrow \rkhstime$ is then
	\begin{equation}
		(\linreg^{*} \tilde{\bm{u}})(\bm{r}) = \mathcal{F}^{-1} \int_{\unitsphere} \bm{E}(\bm{r}, \unitvec{\bm{d}}) \bm{W}(\hat{\bm{d}})^{*} \pwcoeff{\bm{u}}(\hat{\bm{d}}) ds(\hat{\bm{d}}).
	\end{equation}

	The kernel function $\tilde{\Gamma}_r : \Omega \times \Omega \rightarrow \tdmatrix$ of the associated space $\rkhstime_{\linreg}$ is then
	\begin{equation}
		\begin{aligned}
		&\tilde{\Gamma}_r(\bm{r}, \bm{r}') =  \mathcal{F}^{-1} \int_{\unitsphere} \bm{E}(\bm{r}, \unitvec{\bm{d}}) \\
		& \Bigl( \bm{W}(\unitvec{\bm{d}})^{*} \bm{W}(\unitvec{\bm{d}})\Bigr)^{-1} \bm{E}(-\bm{r}', \unitvec{\bm{d}})  ds(\unitvec{\bm{d}})  \mathcal{F}.
		\end{aligned}
	\label{eq:directionally-weighted-solution-basis}
	\end{equation}
	where it is used that $\bm{E}(\bm{r}, \unitvec{\bm{d}})^{*} = \bm{E}(-\bm{r}, \unitvec{\bm{d}})$. The expression corresponds to \eqref{eq:kernel-function-HR} for the specific choice of $\linreg$ defined in \eqref{eq:directional-weighting-operator}. The kernel function can be used in \eqref{eq:optimal-a-regularized} to obtain a sound field estimate. If $\bm{W}$ is chosen to be diagonal for all $\unitvec{\bm{d}}$, the frequency-wise weighting from previous works is recovered. The directional weighting defined here has more degrees of freedom as it also allows the relationship between frequencies to be taken into account. Finding general weighting functions $\bm{W}(\unitvec{\bm{d}})$ from data, and solving the integral in \eqref{eq:directionally-weighted-solution-basis} numerically has been explored for the single-frequency case in \cite{ribeiroSound2024, ribeiroPhysicsconstrained2024}. 
	
	
	The integral in \eqref{eq:directionally-weighted-solution-basis} can be simplified if the directional weighting function is diagonal and each element takes the form $(\bm{W}(\unitvec{\bm{d}}))_{ll} = e^{\frac{\beta_l}{2} \bm{\eta}_l^\top \unitvec{\bm{d}}}$, where $\bm{\eta}_l \in \unitsphere$ represents the direction of the weighting and $\beta_l \in \real_{\geq 0}$ represents the strength of the weighting \cite{uenoDirectionally2021}. This directional weighting is up to a scaling equivalent to the von Mises-Fisher distribution \cite[Section 3.5.4]{mardiaDirectional2000}. With this weighting, estimates where the sound field is primarily propagating in the direction $\bm{\eta}_l$ are preferred. 
	

	The weighting function $\bm{W}$ is then self-adjoint, which leads to $(\bm{W}(\unitvec{\bm{d}})^{*} \bm{W}(\unitvec{\bm{d}}) )_{ll} =  e^{\beta_l \bm{\eta}_l^\top \unitvec{\bm{d}}}$. The strength of the weighting is restricted to be $\beta_l = 0$ for $l=\frac{L}{2}$ to ensure a closed-form solution of \eqref{eq:directionally-weighted-solution-basis}. The regularized kernel function then follows from  the identity \eqref{eq:plane-wave-sinc-identity}, and can be represented by a diagonal matrix defined as 
	\begin{equation}
		(\Gamma_r(\bm{r}, \bm{r}'))_{ll} = \begin{cases}
			\kappa_{l}^{\text{dir}}(\bm{r}, \bm{r}') & 0 \leq l < \frac{L}{2} \\
			\frac{1}{2} (\kappa_{\omega_l}(\bm{r}, \bm{r}')  + \kappa_{\omega_l}(\bm{r}, -\bm{r}') ) & l = \frac{L}{2},
		\end{cases}
		\label{eq:directionally-weighted-diagonal-solution-basis-vonmises}
	\end{equation}
	where the function $\kappa_l^{\text{dir}} : \Omega \times \Omega \rightarrow \complex$ is
	\begin{equation}
			\begin{aligned}
				\kappa_{l}^{\text{dir}}(\bm{r}, \bm{r}') &= j_0\Bigl(\sqrt{\bm{\xi}_{l}^{\top} \bm{\xi}_{l}} \Bigr) \\
				\bm{\xi}_{l} &= \frac{\omega_{l}}{c} (\bm{r} - \bm{r}') - \imag \beta_{l} \bm{\eta}_{l}. 
		\end{aligned}
	\end{equation}

	\subsection{Data weighting envelopes}\label{sec:envelopes}
		\begin{figure}
		\includegraphics{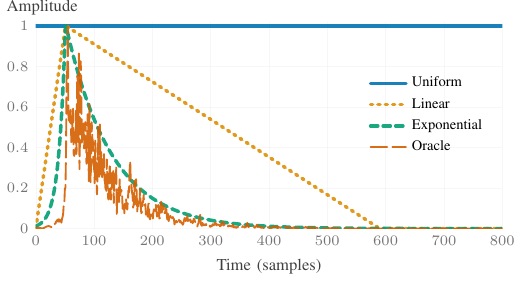}
		\caption{Examples of envelopes used for the data weighting $\tilde{\bm{Q}}$.}
		\label{fig:envelope}
	\end{figure}
	The data weighting operator $\tilde{\bm{Q}}$ should be chosen such that a higher weight is given to data of high quality or importance. The time-domain characteristics of \glspl{rir} are predictable, which can be used to construct such a weighting. \Glspl{rir} generally exhibit an exponential decrease after an initial delay, while noise affecting \glspl{rir} is distributed uniformly in time. Therefore, the \gls{snr} varies drastically over the duration of the \gls{rir}, and the data could be weighted accordingly. 
	
	With this motivation, the data weighting for each microphone $\tilde{\bm{Q}}_m$ is chosen to be diagonal, with diagonal elements $(\tilde{\bm{Q}}_m)_{ll} = (\tilde{\bm{q}}_m)_l$ for $l=0,\dots,L-1$, where the time-domain sequence $\tilde{\bm{q}}_m \in \td$ is referred to as an envelope. The envelope is intended to be an approximation of the magnitude of the \gls{rir}, which then also encodes the \gls{snr} over time. Commonly available prior knowledge that can be used to construct the envelope are acoustic parameters of the space, as well as the position of the loudspeaker and microphones. 
	
	An \textit{individual} version of the envelope can be considered, where $\tilde{\bm{Q}}_m$ is created from information associated with microphone $m$. A \textit{non-individual} version can be considered where aggregate information is used from all microphones, leading to $\tilde{\bm{Q}}_m = \tilde{\bm{Q}}_{m'}$ for all $m, m'$. The former requires more information, but allows for more detailed control of the weighting. 
	
	The inverse data-weighting operator is used in \eqref{eq:optimal-a-regularized}, therefore it must be be invertible, implying that all elements of $\tilde{\bm{q}}$ must be non-zero. It is sufficient to choose each element according to $(\tilde{\bm{Q}}_m)_{ll} = \max(q_{\text{min}}, (\tilde{\bm{q}}_m)_l)$ for some small $q_{\text{min}} \in \real_{> 0}$. 

	Four types of envelopes are considered here, which differ in how accurate they approximate the \gls{rir}, and which prior knowledge is required. Examples of the considered envelopes can be seen in Fig.~\ref{fig:envelope}.

	\subsubsection{Uniform} The envelope is set to $(\tilde{\bm{q}})_l = 1$ for all $l$, corresponding to an identity $\tilde{\bm{Q}}$, whereby all data is weighted equally. The weighting is therefore equivalent to previous methods without data weighting. The \textit{individual} version of this envelope equals the \textit{non-individual} version. 
	
	
	\subsubsection{Exponential} The propagation delay $l_0$ in samples and the reverberation time $\text{RT}_{60}$ is used to model the \gls{rir} \cite{vanwaterschootOptimal2005}. These two pieces of information define the envelope as
	\begin{equation}
		(\tilde{\bm{q}})_l = \begin{cases}
			10^{3 (l-l_0)/(\tau_{\text{init}} f_s)} & l < l_0  \\
			10^{-3 (l-l_0)/(\tau_{\text{decay}} f_s)} & l_0 \leq l.
		\end{cases}
	\end{equation}
	 The initial time-constant $\tau_{\text{init}}$ is meant to capture the fast but non-instantaneous onset of a sampled \gls{rir}, and can be viewed as a parameter to be chosen. The second time-constant $\tau_{\text{decay}}$ is set according to the measured $\text{RT}_{60}$. Note that the delay induced by the recording equipment should be included in the propagation delay. The \textit{non-individual} envelope can make use of the minimum propagation time of all microphones, and the median $\text{RT}_{60}$.

	 

	\subsubsection{Linear} The same information as \textit{exponential} is required, but a linear function is used instead of an exponential function, defined as 
		\begin{equation}
		(\tilde{\bm{q}})_l = \begin{cases}
			\frac{l}{l_0} & l < l_0 \\
			1 - \frac{l - l_0}{f_s \tau_{\text{decay}}} & l_0 \geq l \leq l_0 + f_s \tau_{\text{decay}} \\
			0 & \text{otherwise}.
		\end{cases}
	\end{equation}
	The envelope is expected to perform worse than \textit{exponential}, as it does not capture the decay rate that can be expected of a \gls{rir}.

	\subsubsection{Oracle} The absolute value of the true \glspl{rir} at the microphone positions. The envelope has the correct decay rate like \textit{exponential} while also capturing finer detail of the \gls{rir}. It can be expected to perform better than the other envelopes, while being infeasible to implement in practice. The \textit{non-individual} envelope uses the mean of the absolute values of the \glspl{rir} at the microphones.

	\section{Evaluation}\label{sec:evaluation}

	To demonstrate the effectiveness of the proposed approach, a set of numerical experiments are presented. The evaluation is made by comparing the estimated function $\tilde{\bm{u}}$ with the true function $\tilde{\bm{u}}^{\text{true}}$ on a set of $E$ evaluation points, referred to as $\mathcal{E} = \{\bm{r}_1, \dots, \bm{r}_E\}$. The primary evaluation metric is the \gls{nmse}, defined as
	\begin{equation}
		\text{NMSE} = \frac{\sum_{\bm{r} \in \mathcal{E}} \lVert \tilde{\bm{u}}(\bm{r}) - \tilde{\bm{u}}^{\text{true}}(\bm{r}) \rVert_{\td}^2}{\sum_{\bm{r} \in \mathcal{E}} \lVert \tilde{\bm{u}}^{\text{true}}(\bm{r}) \rVert_{\td}^2} 
	\end{equation}
	Except where explicitly mentioned, all results shown are the mean results from 10 experiments, with random microphone positions and noise realizations. All experiments are performed at a sampling rate of $f_s = \qty{1600}{\hertz}$. The regularization parameter $\lambda$ is chosen as $\frac{\sigma_{p}^2 - \sigma_{s}^2}{10 \sigma_{s}^2}$, where $\sigma_{p}^2$ and $\sigma_{s}^2$ are the measured powers of the microphone signal and the noise signal respectively, equivalent to $\frac{1}{10 \text{SNR}}$. The minimum value for all envelopes is chosen as $q_{\text{min}} = 10^{-6}$, and the initial time-constant for the \textit{exponential} envelope is chosen as $\tau_{\text{init}} = \qty{0.05}{\second}$. 
	

	In the following, the three considered types of sound field data will be described, followed by the four noise models, and finally the evaluation results will be presented. 

	\subsection{Sound field data}
	\begin{figure}
		\includegraphics{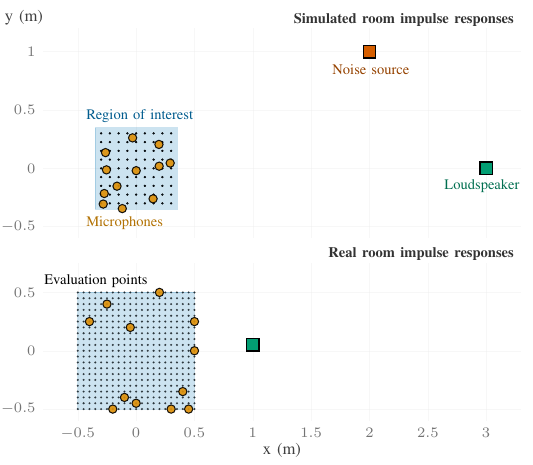}
		\caption{Positions of microphones, loudspeaker and evaluation points for the simulated \glspl{rir} (top) and the real \glspl{rir} (bottom). The microphone positions are sampled randomly within the region of interest, and the figure shows one such realization. }
		\label{fig:stationary-mic-positions}
	\end{figure}
	
	\subsubsection{Free field}
	The \glspl{rir} are generated according to the free field Green's function, using \gls{fir} fractional delay filters with length 81 samples. All sources and microphones are linear and omnidirectional. The length of the sequences are set to $L = 250$, corresponding to \qty{0.16}{\second}. The \glspl{rir} are filtered with a zero-phase highpass filter with a cutoff frequency of \qty{50}{\hertz}, to simulate the lack of low-frequency reproduction from a loudspeaker. The region of interest is a cuboid of size \qtyproduct{70 x 70 x 25}{\centi\meter}. The sound field is evaluated on $243$ evaluation points placed in the region of interest on a uniform grid with a distance of \qty{7.5}{\centi \meter}. The microphone positions are sampled from a uniform distribution within the region of interest. The position of the loudspeaker, microphones, and evaluation points projected onto the same horizontal plane can be seen in Fig.~\ref{fig:stationary-mic-positions}. 
	
	\subsubsection{Simulated room impulse responses}
	\Glspl{rir} from a simulated cuboid room of size \qtyproduct{5.4 x 4.3x 3.2}{\meter} are generated using the image-source method \cite{allenImage1979, scheiblerPyroomacoustics2018}. The room has a reverberation time of approximately $\text{RT}_{60} = \qty{0.36}{\second}$. The geometry of microphone, loudspeaker, and evaluation points is identical to the free field data. The \glspl{rir} are filtered identically to the free field data. The length of the sequences are set to $L = 800$, corresponding to \qty{0.5}{\second}. 
	
	\subsubsection{Real room impulse responses}
	Recorded impulse responses from a real room, taken from the MeshRIR dataset \cite{koyamaMeshRIR2021}, which has a reverberation time of $\text{RT}_{60} = \qty{0.19}{\second}$. From the 441 microphone positions, 12 are chosen at random with uniform distribution to be used as measurements, while the other 429 are used as evaluation points. The positions of the microphones and loudspeaker is shown in Fig.~\ref{fig:stationary-mic-positions}. The length of the sequences are set to $L = 800$, corresponding to \qty{0.5}{\second}

	\subsection{Noise models}
	Four noise models are considered, in order to characterize the impact of different types of noise. 
	
	\subsubsection{Additive white noise} The simplest noise model considered. The data vectors $\tilde{\bm{h}}_m$ are generated according to \eqref{eq:data-model}, with each element of $\tilde{\bm{\epsilon}}$ being zero mean white Gaussian noise. For real data, some level of uncorrelated white noise is present due to thermal noise in the microphones. With high quality microphones, the level of this noise is low. 
	
	\subsubsection{Localized white noise} A spatially correlated noise. The noise is white Gaussian noise propagating from a point source in the room, the position of which is shown in Fig~\ref{fig:stationary-mic-positions}. The measurements are described by \eqref{eq:data-model-sound-pressure}, where the loudspeaker signal $\phi$ is chosen as a perfect sweep \cite{antweilerPerfect1994, antweilerNLMStype2008}, which is particularly suitable for acoustic measurements. The measured signal is deconvolved with a linear operation, leading to data vectors $\tilde{\bm{h}}_m$ consisting of the true \gls{rir} plus an error term dependent on $s$ and $\phi$. 
	
	\subsubsection{Localized pink noise} A noise with considerable temporal correlation, in addition to spatial correlation. The noise is generated in the same way as \textit{localized white noise}, but the signal propagating from the noise source is pink noise. The pink noise is generated by sampling white Gaussian noise, and weighting each frequency by $\frac{1}{\sqrt{f}}$ in the \gls{dft} domain. 
	
	\subsubsection{Wind noise} A temporally non-stationary noise. It is simulated according to \cite{nelkeMeasurement2014}, using the implementation \cite{nelkeWind2014}. Each microphone has an independent realization of wind noise directly added to the measured signal, meaning that the simulated wind noise directly corresponds to $s_m$ in \eqref{eq:data-model-sound-pressure}. The same loudspeaker signal is used as in \textit{localized white noise}. 
	
		\begin{figure}
			\includegraphics{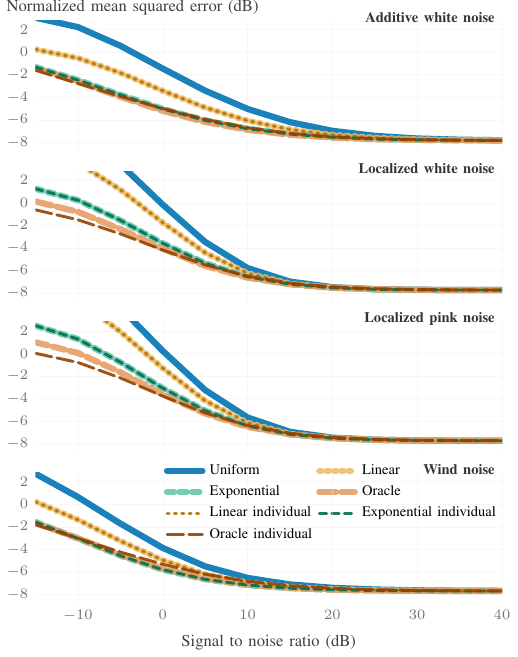}
		\caption{Estimation performance for different noise levels and data weighting envelopes. The noise models are \textit{additive white noise}, \textit{localized white noise}, \textit{localized pink noise}, and \textit{wind noise}, from top to bottom. The data points are at \gls{snr} values -15, -10, \dots, 40 \unit{\decibel}.}
		\label{fig:mse-vs-noise-power}
	\end{figure}

	\subsection{Comparison between data weighting envelopes}
	In this section the envelopes described in Section~\ref{sec:envelopes} are compared, using simulated \glspl{rir}, using all noise models. The \gls{nmse} for all considered envelopes and noise models is shown in Fig.~\ref{fig:mse-vs-noise-power}. The results show that using the data weighting with all considered non-uniform envelopes robustly improves estimation performance when the noise level is significant. When there is little noise, the impact of the data weighting is low or none, but without degrading performance compared to the uniform weighting.
	
	The \textit{oracle} envelope largely performs the best, followed by the exponential envelope, followed by the linear envelope. This suggests that choosing the envelope according to the best available model of the \glspl{rir} is a well-founded strategy. The exponential envelope performs similarly to the unattainable oracle envelope, which suggests that it would be an effective choice in practice. 
	
	\textit{Oracle individual} outperforms \textit{oracle} for the two localized noise models, but not for the spatially uncorrelated noise models. However, the performance of the individually constructed linear and exponential envelopes almost exactly matches their \textit{non-individual} counterparts. That suggests that in a practical scenario, using e.g. \textit{exponential} envelope, the same envelope can be used for all microphones.
	
	\begin{figure}
		\includegraphics{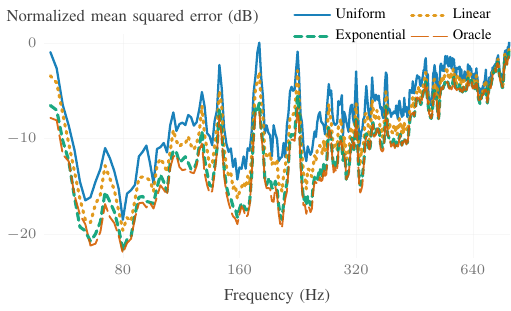}
		\caption{Estimation performance per frequency for different data weighting envelopes, where \textit{additive white noise} is added at a \gls{snr} of \qty{10}{\decibel}. }
		\label{fig:mse-per-freq-envelope}
	\end{figure}
	
	To illustrate the frequency-dependence of the improvement in estimation performance, the \gls{nmse} as a function of frequency is shown in Fig.~\ref{fig:mse-per-freq-envelope} for \textit{additive white noise} at a \gls{snr} of \qty{10}{\decibel}. The figure shows that for a spectrally flat noise, the improvement is also relatively uniformly distributed with regards to frequency. The \textit{individual} envelopes are not shown, as they perform similarly well, as Fig.~\ref{fig:mse-vs-noise-power} shows.

	\begin{figure}
		\includegraphics{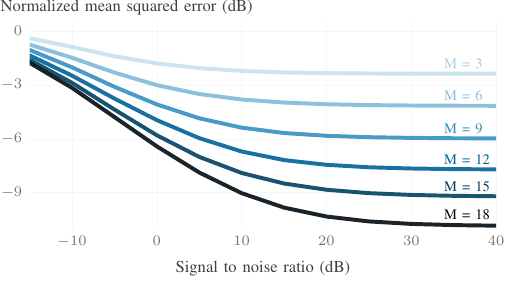}
		\caption{Estimation performance for different number of microphones $M$, and levels of \textit{additive white noise}. The \textit{exponential} data weighted estimator is used. The data points are at \gls{snr} values -15, -10, \dots, 40 \unit{\decibel}.}
		\label{fig:mse-vs-snr-vs-mics}
	\end{figure}

	The same experiment is performed for \textit{additive white noise} with the \textit{exponential} envelope for different numbers of microphones. In Fig.~\ref{fig:mse-vs-snr-vs-mics} the \gls{nmse} is shown for between $3$ and $18$ microphones. As expected, the \gls{nmse} decreases as the number of microphones increase, but the difference is much greater at high \glspl{snr}. In addition, the noise level is comparatively less impactful when the number of microphones is low, due to the spatial undersampling being the dominant source of error.

	\subsection{Directional weighting combined with data weighting}
	The directional weighting as described in Section~\ref{sec:directional-regularization} can be combined with the data weighting, which is investigated in this section. The free field and simulated \glspl{rir} data is used, together with \textit{additive white noise}. Only the exponential envelope is considered, as it performs well while still being practical. The directional weighting \eqref{eq:directionally-weighted-diagonal-solution-basis-vonmises} is used, which can be computed in closed form. The method with directional weighting is referred to as \textit{directional}, while the non-weighted method, i.e. with identity $\linreg$, is referred to as \textit{diffuse}. The chosen $\unitvec{\bm{\eta}}$ is the direction from the loudspeaker towards the center of the region of interest. For an investigation of more flexible directional weighting functions, the reader is referred to \cite{ribeiroSound2024}. 
	
		\begin{figure}
			\includegraphics{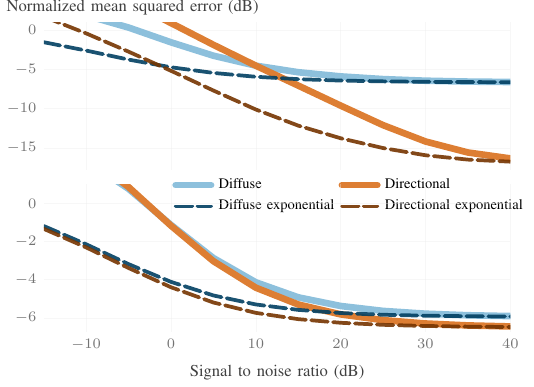}
			\caption{Estimation performance in free field (top) and reverberant (bottom) conditions, for different levels of \textit{additive white noise}. The data points are at \gls{snr} values -15, -10, \dots, 40 \unit{\decibel}.}
			\label{fig:envelope-with-dir-mse-vs-snr}
		\end{figure}
	
	In free field, due to a lack of reflections the entire sound field is propagating in approximately the same direction. This scenario is close to ideal conditions for the specific directional weighting used, and therefore a large weighting parameter of $\beta_l = 5$ is used for all $l$ except $\beta_{\frac{L}{2}} = 0$. In free field it is not reasonable to use the true value for the $\text{RT}_{60}$ in order to construct the time-domain data weighting, as there is no reverberation, and therefore no well-defined reverberation time. Instead, a low but non-zero value is used, chosen as $\tau_{\text{decay}} = \qty{0.05}{\second}$. 
	
	

\begin{figure}
		\includegraphics{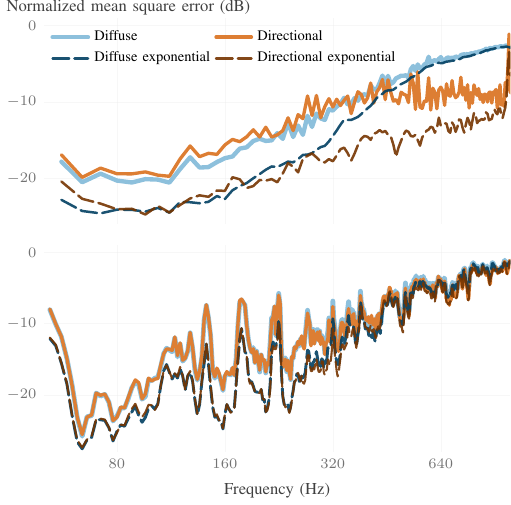}
		\caption{Estimation performance per frequency in free field (top) and using simulated \glspl{rir} (bottom), at an \gls{snr} of \qty{20}{\decibel}.}
		\label{fig:mse-per-freq-envelope-with-dir}
	\end{figure}

	The \gls{nmse} for different levels of \textit{additive white noise} is shown in Fig.~\ref{fig:envelope-with-dir-mse-vs-snr}. The directional weighting dramatically increases estimation performance at high \gls{snr}, but degrades fast as the noise level increases. Combining the directional and exponential weighting leads to high estimation performance even at lower \glspl{snr}. The \gls{nmse} per frequency is shown in Fig.~\ref{fig:mse-per-freq-envelope-with-dir}. It can be seen that the directional weighting improves the estimation performance, in particular for higher frequencies.  The time-domain data weighting improves the estimation performance in particular for low frequencies. By combining the directional and data weighting, the estimate is improved over all frequencies, and in the high frequencies it is even better than only the directional weighting. 

	In the reverberant environment, less of the sound field is propagating in the same direction compared to the free field scenario, so a smaller weighting parameter of $\beta_l = 1$ is chosen.  The \gls{nmse} per \gls{snr} is shown in Fig.~\ref{fig:envelope-with-dir-mse-vs-snr}, which displays a similar pattern as in free field, but with a smaller improvement provided by the directional weighting. The \gls{nmse} per frequency is shown in Fig.~\ref{fig:mse-per-freq-envelope-with-dir}. The data weighting improves the performance both when the directional weighting is in use and when it is not. A summary of the total \gls{nmse} for all considered methods for both free field and reverberant conditions is shown in Table~\ref{table:nmse-envelope-with-dir}, where it can be seen that the combined directional weighting and data weighting provides the best estimation performance for both scenarios. 

		\begin{table}
		\centering
		\caption{Normalized mean square error in decibels for simulated \glspl{rir} with additive white noise at a \gls{snr} of \qty{20}{\decibel}.}
		\label{table:nmse-envelope-with-dir}
		\begin{tabular}{l S[detect-weight, table-format=4.2, round-mode=places, round-precision=2] 
				S[detect-weight, table-format=4.2, round-mode=places, round-precision=2] S[detect-weight, table-format=4.2, round-mode=places, round-precision=2] 
				S[detect-weight, table-format=4.2, round-mode=places, round-precision=2]}
			\toprule 
			& \multicolumn{2}{c}{Free field} & \multicolumn{2}{c}{Reverberant} \\
			& {Uniform} & {Exponential} & {Uniform} & {Exponential} \\
			\midrule 
			Diffuse &  -5.899367311749982  & -6.429843896497674 &  -5.3833320723662395  & -5.752071719076283 \\
			Directional & -9.618767070822397  & \bfseries -13.776593745626933  & -5.823790994266678  & \bfseries  -6.264756470234397 \\
			\bottomrule
		\end{tabular}
	\end{table}

	\subsection{Estimation of real room impulse responses}\label{sec:real-rir-experiment}
	
\begin{figure}
		\includegraphics{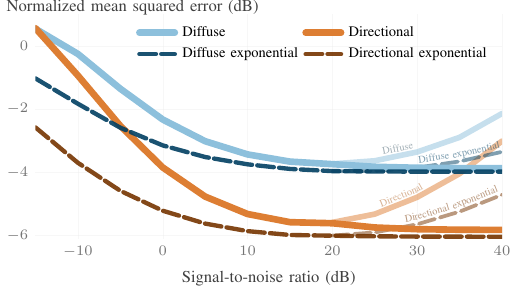}
		\caption{Estimation performance on real \glspl{rir} for different amounts of wind noise. The translucent lines show how the estimation performance degrades when the regularization parameter becomes too small at high \gls{snr} values. The data points are at \gls{snr} values -15, -10, \dots, 40 \unit{\decibel}. }
		\label{fig:meshrir-mse-vs-snr}
	\end{figure}

	Finally, the proposed method is evaluated on the real \gls{rir} data, with \textit{wind noise} as the noise model. The \gls{nmse} as a function of the \gls{snr} is shown in Fig.~\ref{fig:meshrir-mse-vs-snr}, demonstrating the improvement provided by the data weighting. Choosing the regularization parameter according to $\lambda = \frac{1}{10 \text{SNR}}$ as is done in previous experiments causes the error to increase for \glspl{snr} above \qty{20}{\decibel}, which is shown by the translucent lines in Fig.~\ref{fig:meshrir-mse-vs-snr}. The estimators with \textit{exponential} weighting can be noted to be less sensitive to the regularization parameter being chosen too low. To avoid this performance degradation, the regularization is chosen as $\lambda = \max (10^{-3}, \frac{1}{10 \text{SNR}})$, with which the \gls{nmse} continues to decrease as the \gls{snr} increases. In contrast to the simulated \glspl{rir}, the proposed data-weighted estimators outperforms the non-weighted estimators even at the highest \glspl{snr}. The increase in \gls{nmse} for high \gls{snr}, along with the fact that the data weighting outperforms the non-weighted estimates even at very high \glspl{snr} suggests that there is some noise in the dataset either in the \glspl{rir} or microphone positions. 
	
\begin{figure}
		\includegraphics{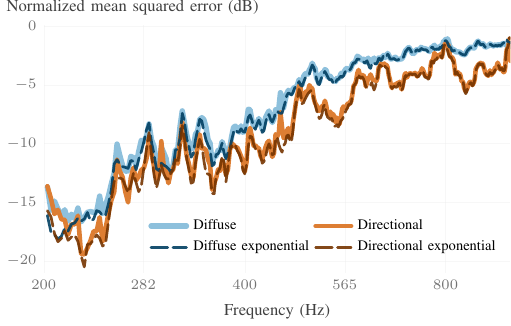}
		\caption{Estimation performance per frequency using real data from \cite{koyamaMeshRIR2021} without added noise.}
		\label{fig:meshrir-mse-per-freq}
	\end{figure}

	In Fig.~\ref{fig:meshrir-mse-per-freq} the \gls{nmse} as a function of frequency is shown, where no noise is added. The \textit{exponential} weighting reduces the error by a smaller amount over the entire frequency range. The mean \gls{nmse} over all frequencies is shown in Table~\ref{table:mse-meshrir}, showing the lowest error for \textit{directional exponential}. This demonstrates that even in very favourable conditions, such as the where the dataset \cite{koyamaMeshRIR2021} was recorded, there is still a benefit in applying the data weighting. 
	
	\begin{table}
	\centering
	\caption{Normalized mean square error in decibels using real data, without any added noise.}
	\label{table:mse-meshrir}
	\begin{tabular}{l S[detect-weight, table-format=4.2, round-mode=places, round-precision=2] 
			S[detect-weight, table-format=4.2, round-mode=places, round-precision=2]}
		\toprule 
		& {Uniform} & {Exponential} \\
		\midrule 
		Diffuse &  -3.8743627229942583  & -3.982792435533722 \\
		Directional & -5.8287453466728785  & \bfseries -6.040884549256051 \\
		\bottomrule
	\end{tabular}
\end{table}

\begin{figure*}
			\includegraphics{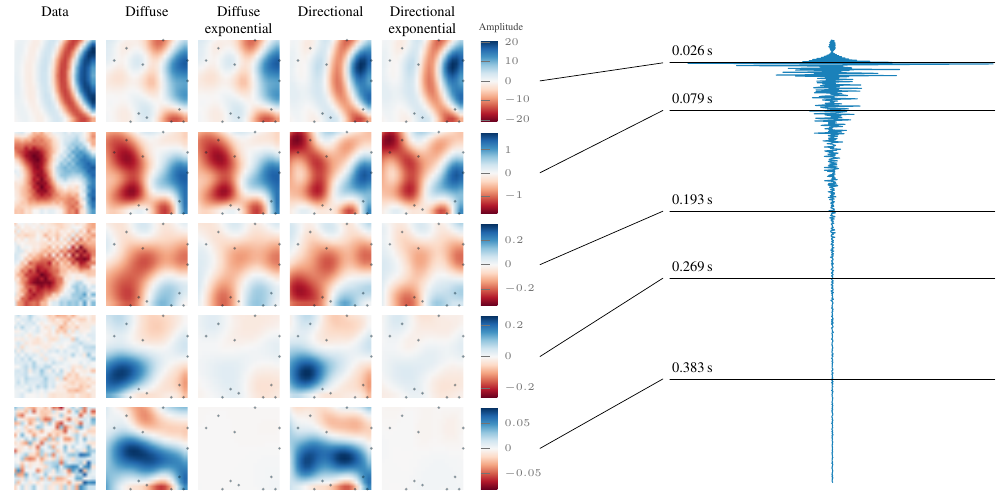}
	\caption{The sound field estimates using the real \gls{rir} data without added noise, shown at all evaluation points, the geometry of which can be seen in Fig.~\ref{fig:stationary-mic-positions}. Each row displays a different moment in time, from early at the top, to late at the bottom. Note that each row has a different scaling. The \gls{rir} shown on the right is taken from \textit{data} at the center of the region. The grey dots shows the microphone positions. }
	\label{fig:meshrir-soundfield-grid}
\end{figure*}

	The estimates over the region of interest can be compared with the data from the dataset for different time indices of the \gls{rir}, shown in Fig.~\ref{fig:meshrir-soundfield-grid}. The estimation of the early part of the \gls{rir} is clearly improved for \textit{directional} and \textit{directional exponential}, by taking the direction of the initial wave into account. However, for later parts of the \gls{rir}, the noise in the data becomes more significant compared to the amplitude of the \gls{rir}, and there the \textit{diffuse exponential} and \textit{directional exponential} performs better, as the non-weighted methods attempt to fit an estimate to mostly noise.

	\section{Conclusion}\label{sec:conclusion}
	A sound field estimation method has been presented, with which time-domain sound fields can be estimated using kernel ridge regression. The method is a generalization of previous kernel ridge regression methods, allowing for the inclusion of prior knowledge relating to time-domain properties of the sound field. A data weighting has been proposed, which can improve estimation performance especially in noisy conditions, by considering the time-domain characteristics of the \gls{rir} as prior knowledge. The weighting has been shown to be effectively modelled by only the propagation delay and reverberation time, which is information that can be obtained in practice. In addition, the data weighting has been shown to be even more effective when combined with a directional weighting, which exploits spatial characteristics of the sound field. 
	
	
	\appendices

	\section{Discrete Fourier transform}\label{sec:discrete-fourier-transform}
	A linear operator $\bm{B} \in \tdmatrix$ can always be represented by a real matrix $\bm{B} \in \real^{L \times L}$. Its adjoint is then represented by $\bm{B}^\top$, which can be shown by
	\begin{equation}
		\begin{aligned}
			\langle \bm{B} \bm{a}, \bm{a}' \rangle_{\td} &= \bm{a'}^\top \bm{B} \bm{a} = \bm{a'}^\top (\bm{B}^\top)^\top \bm{a} = \langle \bm{a}, \bm{B}^\top \bm{a}' \rangle_{\td}. 
		\end{aligned}
	\end{equation}

	Despite having complex-valued vectors, $\fd$ is a real vector space, as can be seen from its inner product \eqref{eq:freq-domain-inner-product}. Any $\bm{a} \in \fd$ can be written in terms of its real and complex elements, which for even $L$ looks like
	\begin{equation}
		\bm{a} = (a_0, a_1 + \imag b_1, \dots, a_{\frac{L}{2}-1} + \imag b_{\frac{L}{2}-1}, a_{\frac{L}{2}}).
	\end{equation}
	The definition is the same for odd $L$, but then $a_{\frac{L}{2}}$ is excluded. An isomorphism $\mathcal{S} : \fd \rightarrow \real^L$ can be defined for even $L$ as
	\begin{equation}
		\mathcal{S} \bm{a} = (a_0, \dots, a_{\frac{L}{2}}, b_1, \dots, b_{\frac{L}{2}-1}) \in \real^L,
	\end{equation}
	where the inverse $\mathcal{S}^{-1} : \real^L \rightarrow \fd$, and the analogous definitions for odd $L$ follow naturally. A linear operator $\bm{A} \in \fdmatrix$ can always be represented by the matrix $\bm{D} \in \real^{L \times L}$, applied to $\bm{a} \in \fd$ as 
 	\begin{equation}
 		\bm{A} \bm{a} = \mathcal{S}^{-1} \bm{D} \mathcal{S} \bm{a}.
 	\end{equation}
	
	In contrast, all such linear operators cannot be represented by complex matrices $\bm{G} \in \complex^{\rdftlen \times \rdftlen}$. Some can be represented in that form, and because of its convenience it is the form primarily used in this paper. In order to guarantee that $(\bm{A} \bm{v})_{0}$ and $(\bm{A} \bm{v})_{\frac{L}{2}}$ are still real-valued, the associated matrix for even and odd $L$ should respectively have the form
	\begin{equation}
		\bm{A}_{\text{even}} =  \begin{bmatrix}
			a & 0 & b \\
			0 & \bm{G}_{\text{even}} & 0 \\
			c & 0& d \\
		\end{bmatrix}, \qquad \bm{A}_{\text{odd}} =  \begin{bmatrix}
		a & 0 \\
		0 & \bm{G}_{\text{odd}} \\
	\end{bmatrix},
	\end{equation}
	for $a,b,c,d \in \real$ and $\bm{G}_{\text{even}} \in \complex^{(\rdftlen-2) \times (\rdftlen-2)}$, $\bm{G}_{\text{odd}} \in \complex^{(\rdftlen-1) \times (\rdftlen-1)}$. The adjoint $\bm{A}^{*}$ is then represented by the matrix $\bm{C}^{-1} \bm{A}^\herm \bm{C}$, as shown by
	\begin{equation}
		\begin{aligned}
			\langle \bm{A} \bm{a}, \bm{b} \rangle_{\fd} &= \mathfrak{Re}[\bm{b}^{\herm} \bm{C} \bm{A} \bm{a}] \\
			&= \mathfrak{Re}[\bm{b}^{\herm} (\bm{C}^{-1} \bm{A}^\herm \bm{C})^{\herm} \bm{C} \bm{a}] \\
			&= \langle \bm{a}, \bm{C}^{-1} \bm{A}^\herm \bm{C} \bm{b} \rangle_{\fd}.
		\end{aligned}
		\label{eq:freq-domain-adjoint}
	\end{equation}

	\section{Frequency-domain reproducing kernel}\label{sec:frequency-domain-rkhs-details}
	The point evaluation operator on $\rkhs$ is denoted by $\mathcal{M}_{\bm{r}} : \rkhs \rightarrow \fd$, and is defined by $\mathcal{M}_{\bm{r}} \bm{u} = \bm{u}(\bm{r})$. The reproducing kernel is characterized by the adjoint of the point evaluation operator $\evalop_{\bm{r}}^{*} : \fd \rightarrow \rkhs$ as \cite[Example 1]{diwaleGeneralized2018}
	\begin{equation}
		\evalop_{\bm{r}}^{*} = \Gamma(\cdot, \bm{r}).
		\label{eq:kernel-freq-function-adjoint-relationship}
	\end{equation}
	A direct consequence of \eqref{eq:kernel-freq-function-adjoint-relationship}, following from the definition of the adjoint $\langle \evalop_{\bm{r}} \bm{u}, \bm{a} \rangle_{\fd} = \langle \bm{u}, \evalop_{\bm{r}}^{*} \bm{a} \rangle_{\rkhs} $ \cite[Ch.2 Theorem 2.2]{conwayCourse1990}, is the well known reproducing property 
	\begin{equation}
		\begin{aligned}
			&\langle \bm{u}(\bm{r}), \bm{a} \rangle_{\fd} = \langle \bm{u}, \Gamma(\cdot, \bm{r}) \bm{a} \rangle_{\rkhs}
		\end{aligned}
		\label{eq:reproducing-property-freq}
	\end{equation}
	for all $\bm{u} \in \rkhs$ and $\bm{a} \in \fd$. 
	
	The left-hand side of \eqref{eq:reproducing-property-freq} can be expanded using \eqref{eq:frequency-domain-plane-wave-decomposition} as
	\begin{equation}
		\begin{aligned}
			\langle \bm{u}(\bm{r}), \bm{a} \rangle_{\fd} &= \int_{\unitsphere}  \langle  \bm{E}(\bm{r}, \unitvec{\bm{d}}) \pwcoeff{\bm{u}}(\unitvec{\bm{d}}), \bm{a} \rangle_{\fd} \,ds(\unitvec{\bm{d}})\\
			&= \int_{\unitsphere}  \langle \pwcoeff{\bm{u}}(\unitvec{\bm{d}}),  \bm{E}(-\bm{r}, \unitvec{\bm{d}}) \bm{a} \rangle_{\fd} \,ds(\unitvec{\bm{d}}).
		\end{aligned}
	\label{eq:appendix-kernel-freq-ip}
	\end{equation}
	From the inner product definition in \eqref{eq:multi-freq-rkhs-inner-product}, \eqref{eq:appendix-kernel-freq-ip} can be identified as the inner product between $\bm{u}$ and a function in $\rkhs$ defined by the directionality function $\bm{E}(-\bm{r}, \unitvec{\bm{d}}) \bm{a}$. Inserting the directionality function into \eqref{eq:frequency-domain-plane-wave-decomposition} gives
	\begin{equation}
		\begin{aligned}
			\Gamma(\bm{r}, \bm{r}')\bm{a} &= \int_{\unitsphere} \bm{E}(\bm{r}, \unitvec{\bm{d}}) \bm{E}(-\bm{r}', \unitvec{\bm{d}}) \,ds(\unitvec{\bm{d}}) \bm{a},  
		\end{aligned}
		\label{eq:freq-domain-kernel-derivation}
	\end{equation}
	for which the integrand is 
	\begin{equation}
		\begin{aligned}
		&(\bm{E}(\bm{r}, \unitvec{\bm{d}}) \bm{E}(-\bm{r}', \unitvec{\bm{d}}))_{ll} \\
		&= \begin{cases}
			e^{-\imag\frac{\omega_l}{c} (\bm{r} - \bm{r}')^\top \unitvec{\bm{d}}} & 0 \leq l < \frac{L}{2}\\
			\frac{1}{2}(\cos(\frac{\omega_l}{c} (\bm{r} - \bm{r}')^\top \unitvec{\bm{d}}) \\ \quad + \cos(\frac{\omega_l}{c} (\bm{r} + \bm{r}')^{\top} \unitvec{\bm{d}})) & l = \frac{L}{2}.
		\end{cases}
	\end{aligned}
	\end{equation}
	The expression can be simplified using the identity \cite[Eq.65]{uenoDirectionally2021}
	\begin{equation}
		\begin{aligned}
			\int_{\unitsphere} e^{-\imag \bm{r}^{\top} \unitvec{\bm{d}}} \;ds(\unitvec{\bm{d}}) &= j_0(\sqrt{ \bm{r}^{\top} \bm{r}}),
		\end{aligned}
	\label{eq:plane-wave-sinc-identity}
	\end{equation}
	which holds for both real and complex $\bm{r}$. Note that $\sqrt{ \bm{r}^{\top} \bm{r}}$ is identical to the Euclidian norm when $\bm{r}$ is real, but not when it is complex. For real $\bm{r}$, also $\int_{\unitsphere} \cos(\bm{r}^{\top} \unitvec{\bm{d}}) \;ds(\unitvec{\bm{d}}) = j_0(\sqrt{ \bm{r}^{\top} \bm{r}})$ holds. The kernel function is then
		\begin{equation}
		\begin{aligned}
			(\Gamma(\bm{r}, \bm{r}'))_{ll} &= 
			\begin{cases}
				j_0(\frac{\omega_{l}}{c} \lVert \bm{r} - \bm{r}' \rVert_2) & 0 \leq l < \frac{L}{2} \\
				 \frac{1}{2} (j_0(\frac{\omega_{l}}{c} \lVert \bm{r} - \bm{r}' \rVert_2) \\ 
				 \quad + j_0(\frac{\omega_{l}}{c} \lVert \bm{r} + \bm{r}' \rVert_2)) & l = \frac{L}{2}.
			\end{cases}
		\end{aligned}
	\end{equation}
	

	\section{Time-domain reproducing kernel}\label{sec:time-domain-rkhs-details}
	The time-domain kernel can be derived similarly to the frequency-domain kernel, using the reproducing property $\langle \tilde{\bm{u}}(\bm{r}), \tilde{\bm{a}} \rangle_{\td} = \langle \tilde{\bm{u}}, \tilde{\Gamma}(\cdot, \bm{r}) \tilde{\bm{a}} \rangle_{\rkhstime}$ for all $\tilde{\bm{a}} \in \td$ and $\tilde{\bm{u}} \in \rkhstime$. 

	\begin{equation}
	\begin{aligned}
		\langle \tilde{\bm{u}}(\bm{r}), \tilde{\bm{a}} \rangle_{\td} &=  \int_{\unitsphere} \langle \mathcal{F}^{-1} \bm{E}(\bm{r}, \unitvec{\bm{d}}) \pwcoeff{\bm{u}}(\unitvec{\bm{d}}), \tilde{\bm{a}} \rangle_{\td}  \,ds(\unitvec{\bm{d}})  \\
		&= \int_{\unitsphere} \langle \pwcoeff{\bm{u}}(\unitvec{\bm{d}}), \bm{E}(-\bm{r}, \unitvec{\bm{d}}) \mathcal{F} \tilde{\bm{a}} \rangle_{\fd}  \,ds(\unitvec{\bm{d}})  \\
		&= \langle \tilde{\bm{u}}, \tilde{\Gamma}(\cdot, \bm{r}) \tilde{\bm{a}} \rangle_{\rkhstime}.
	\end{aligned}
\label{eq:kernel-derivation-time-domain}
\end{equation}
The last equality allows the directionality function of $\tilde{\Gamma}(\cdot, \bm{r}) \tilde{\bm{a}}$ to be identified as $\bm{E}(-\bm{r}, \unitvec{\bm{d}}) \mathcal{F} \tilde{\bm{a}}$, which can be inserted into \eqref{eq:time-domain-sound-field-function} and \eqref{eq:herglotz} to obtain
	\begin{equation}
	\begin{aligned}
		 \tilde{\Gamma}(\bm{r}, \bm{r}') \tilde{\bm{a}} &= \mathcal{F}^{-1} \int_{\unitsphere}  \bm{E}(\bm{r}, \unitvec{\bm{d}})  \bm{E}(-\bm{r}', \unitvec{\bm{d}}) \,ds(\unitvec{\bm{d}}) \mathcal{F} \tilde{\bm{a}} \\
		 &= \mathcal{F}^{-1}\Gamma(\bm{r}, \bm{r}') \mathcal{F} \tilde{\bm{a}},
	\end{aligned}
\end{equation}
from which the kernel function can be easily identified.


	\section{Equivalence between time- and frequency-domain kernel ridge regression}\label{sec:time-frequency-equivalence}
	The proposed method with identity weighting-operators $\linreg$ and $\tilde{\bm{Q}}$ is equivalent to a previously proposed frequency-domain \gls{krr} approach \cite{uenoKernel2018}, which will be shown in this section. The equivalence holds except for the Nyquist frequency, which is not explicitly modelled in \cite{uenoKernel2018} as it is derived for continuous frequencies $\omega$. The two methods will therefore give different estimates for the Nyquist frequency. Hence, the following derivation will be made assuming odd $L$ where there is no frequency bin at index $\frac{L}{2}$. In practice, the method in \cite{uenoKernel2018} is applied to data obtained using the \gls{dft}, which means that the data at the Nyquist frequency will also be real-valued, even though this is not explicitly modelled. 
	
	
	The frequency-domain estimate for frequency $\omega$ as obtained by \gls{krr} in \cite[Eq. 24]{uenoKernel2018} is given by  
	\begin{equation}
		\begin{aligned}
		p(\bm{r}, \omega) &= \bm{\kappa}(\bm{r}, \omega)^\top (\bm{K}(\omega) + \lambda \bm{I})^{-1} \bm{p} (\omega),
		\end{aligned}
	\label{eq:single-freq-solution}
	\end{equation}
	where $\bm{p}(\omega) \in \complex^M$ represents the data for all microphones for frequency $\omega$, the system matrix $\bm{K}(\omega) \in \complex^{M \times M}$ is defined as $(\bm{K}(\omega))_{mm'} = \kappa_{\omega}(\bm{r}_m, \bm{r}_{m'})$, and $\bm{\kappa}(\bm{r}, \omega) \in \complex^{M}$ is defined as $(\bm{\kappa}(\bm{r}, \omega))_{m} = \kappa_{\omega}(\bm{r}, \bm{r}_{m})$.
%

	Concatenating the variables for each considered frequency, assumed to be the frequencies sampled by the \gls{dft}, the estimate $\bm{p}(\bm{r}) \in \complex^{\rdftlen}$ is
	\begin{equation}
		\begin{aligned}
		\bm{p}(\bm{r}) &= \bm{\kappa}(\bm{r})^{\top}(\bm{K} +  \lambda \bm{I})^{-1} \bm{p},
		\end{aligned}
		\label{eq:krr-freq-domain-solution}
	\end{equation}
	where the constituent variables are
		\begin{equation}
		\begin{aligned}
			\bm{p} &= (\bm{p}(\omega_0), \dots,  \bm{p}(\omega_{\rdftlen-1})  \in \complex^{M\rdftlen} \\
			\bm{K} &= \diag\{\bm{K}(\omega_l)\}_{l=0}^{\rdftlen-1} \in \complex^{M\rdftlen \times M\rdftlen }\\
			\bm{\kappa}(\bm{r}) &= \diag\{\bm{\kappa}(\bm{r}, \omega_l) \}_{l=0}^{\rdftlen-1} \in \complex^{\rdftlen \times M\rdftlen}.
		\end{aligned}
		\label{eq:frequency-stacked-variables}
	\end{equation}
	

	
	For comparison, the time-domain estimate shown in \eqref{eq:optimal-form-stationary-microphones} can be expressed as 
	\begin{equation}
		\tilde{\bm{u}}(\bm{r}) = \tilde{\Gamma}(\bm{r}) (\tilde{\Gamma} + \lambda \bm{I})^{-1} \tilde{\bm{h}}, 
		\label{eq:time-domain-reconstruction-matrix}
	\end{equation}
	where $\tilde{\Gamma}(\bm{r}) :  \td^M \rightarrow \td$ can be represented by the matrix
	\begin{equation}
		\tilde{\Gamma}(\bm{r}) = \begin{bmatrix}
			\tilde{\Gamma}(\bm{r}, \bm{r}_1) & \dots & \tilde{\Gamma}(\bm{r}, \bm{r}_M)
		\end{bmatrix}. 
	\end{equation}
	The time-domain data vector and the frequency-domain data vector can be related by $\tilde{\bm{h}} = \mathcal{F}_M^{-1} \bm{P} \bm{p}$, where $\mathcal{F}_M^{-1} = \diag\{\mathcal{F}^{-1}\}_{m=1}^{M} : \fd^M \rightarrow \td^M$, and $\bm{P} : \complex^{M\rdftlen} \rightarrow \fd^M$ is a permutation that changes frequency-major order to microphone-major order, after which the vector can straightforwardly be identified with vectors from $\fd^M$. The same constructions can be used for the other quantities in \eqref{eq:krr-freq-domain-solution} and \eqref{eq:time-domain-reconstruction-matrix}, which gives the relationships $\tilde{\Gamma} = \mathcal{F}^{-1}_M \bm{P} \bm{K} \bm{P}^{-1} \mathcal{F}_M$ and $\tilde{\Gamma}(\bm{r}) = \mathcal{F}^{-1} \bm{\kappa}(\bm{r})^\top \bm{P}^{-1} \mathcal{F}_M$. Inserting the relationships into \eqref{eq:time-domain-reconstruction-matrix} leads to the expression
	\begin{equation}
		\begin{aligned}
			\tilde{\bm{u}}(\bm{r})  &= \mathcal{F}^{-1} \bm{\kappa}(\bm{r})^\top \bm{P}^{-1} \mathcal{F}_M (\mathcal{F}^{-1}_M \bm{P} \bm{K} \bm{P}^{-1} \mathcal{F}_M \\
			&+  \lambda \bm{I})^{-1}  \mathcal{F}^{-1}_M \bm{P} \bm{p} \\
			&= \mathcal{F}^{-1} \bm{\kappa}(\bm{r})^\top (\bm{K} +  \lambda\bm{I})^{-1} \bm{p} \\
			&= \mathcal{F}^{-1} \bm{p}(\bm{r}),
		\end{aligned}
		\label{eq:freq-time-relationship}
	\end{equation}
	which is the expected result. This shows that the time-domain solution is the inverse \gls{dft} of the frequency-domain solution if the frequency-domain data is obtained from the \gls{dft}.

	\section{Regularized kernel function}\label{sec:regularized-kernel-function}
	The regularized \gls{krr} in \eqref{eq:krr-stationary-microphones-regularized} is an optimization problem over the space $\rkhstime_{\linreg}$, which has a different inner product and kernel function compared to $\rkhstime$. The reproducing property is $\langle \tilde{\bm{u}}(\bm{r}), \tilde{\bm{a}} \rangle_{\td} = \langle \tilde{\bm{u}}, \tilde{\Gamma}_r(\cdot, \bm{r}) \tilde{\bm{a}} \rangle_{\rkhstime_{\linreg}}$, where the right-hand side can be expanded as  $\langle \tilde{\bm{u}}, \tilde{\Gamma}_r(\cdot, \bm{r}) \tilde{\bm{a}} \rangle_{\rkhstime_{\linreg}} = \langle \linreg \tilde{\bm{u}}, \linreg \tilde{\Gamma}_r(\cdot, \bm{r}) \tilde{\bm{a}} \rangle_{\mathcal{Z}} = \langle \tilde{\bm{u}}, \linreg^{*} \linreg \tilde{\Gamma}_r(\cdot, \bm{r}) \tilde{\bm{a}} \rangle_{\rkhstime}$. Expanding the left-hand side gives
	\begin{equation}
		\begin{aligned}
			\langle \tilde{\bm{u}}(\bm{r}), \tilde{\bm{a}} \rangle_{\td} 
			&= \int_{\unitsphere} \langle \pwcoeff{\bm{u}}(\unitvec{\bm{d}}), \bm{E}(-\bm{r}, \unitvec{\bm{d}}) \mathcal{F} \tilde{\bm{a}} \rangle_{\fd}  \,ds(\unitvec{\bm{d}})  \\
			&= \langle \tilde{\bm{u}}, \linreg^{*} \linreg \tilde{\Gamma}_r(\cdot, \bm{r}) \tilde{\bm{a}} \rangle_{\rkhstime}.
		\end{aligned}
	\end{equation}
	The last equality allows the directionality function of $\linreg^{*} \linreg \tilde{\Gamma}_r(\cdot, \bm{r}) \tilde{\bm{a}}$  to be identified as $\bm{E}(-\bm{r}, \unitvec{\bm{d}}) \mathcal{F} \tilde{\bm{a}}$. This directionality function is identical to the one considered in \eqref{eq:kernel-derivation-time-domain}, which was determined to correspond to $\tilde{\Gamma}(\cdot, \bm{r}') \tilde{\bm{c}}$, which leads to the equality $\linreg^{*} \linreg \tilde{\Gamma}_r(\cdot, \bm{r}) \tilde{\bm{c}} = \tilde{\Gamma}(\cdot, \bm{r}') \tilde{\bm{c}}$. With the assumption that the operator $\linreg$ is invertible, the equality can be rewritten
		\begin{equation}
		\begin{aligned}
		\tilde{\Gamma}_r(\cdot, \bm{r}) \tilde{\bm{a}} &= (\linreg^{*} \linreg)^{-1} \tilde{\Gamma}(\cdot, \bm{r}') \tilde{\bm{a}} ,
			\end{aligned}
	\end{equation}
	which sufficiently defines the regularized kernel. 
	

	\bibliographystyle{IEEEbib_mod}
	\bibliography{abbrev, refs}

\begin{thebibliography}{10}

\bibitem{betlehemPersonal2015}
T.~Betlehem, W.~Zhang, M.~A. Poletti, and T.~D. Abhayapala,
\newblock ``Personal sound zones: Delivering interface-free audio to multiple
  listeners,''
\newblock {\em IEEE Signal Process. Mag.}, vol. 32, no. 2, pp. 81--91, Mar.
  2015.

\bibitem{zhangSurround2017}
W.~Zhang, P.~N. Samarasinghe, H.~Chen, and T.~D. Abhayapala,
\newblock ``Surround by sound: A review of spatial audio recording and
  reproduction,''
\newblock {\em Appl. Sci.}, vol. 7, no. 5, May 2017,
\newblock {A}rt. no. 532.

\bibitem{ajdlerPlenacoustic2006}
T.~Ajdler, L.~Sbaiz, and M.~Vetterli,
\newblock ``The plenacoustic function and its sampling,''
\newblock {\em IEEE Trans. Signal Process.}, vol. 54, no. 10, pp. 3790--3804,
  Oct. 2006.

\bibitem{vanwaterschootOptimally2008}
T.~{van Waterschoot}, G.~Rombouts, and M.~Moonen,
\newblock ``Optimally regularized adaptive filtering algorithms for room
  acoustic signal enhancement,''
\newblock {\em Signal Process.}, vol. 88, no. 3, pp. 594--611, Mar. 2008.

\bibitem{bernschutzSound2012}
B.~Bernsch{\"u}tz, P.~Stade, and M.~R{\"u}hl,
\newblock ``Sound field analysis in room acoustics,''
\newblock in {\em Proc. VDT Int. Conv.}, 2012, pp. 1--22.

\bibitem{florencioMaximum2015}
D.~Florencio and Z.~Zhang,
\newblock ``Maximum a posteriori estimation of room impulse responses,''
\newblock in {\em Proc. {IEEE} Int. Conf. Acoust., Speech, Signal Process.
  ({ICASSP})}, Apr. 2015, pp. 728--732.

\bibitem{jalmbyLowrank2021}
M.~J{\"a}lmby, F.~Elvander, and T.~{van Waterschoot},
\newblock ``Low-rank tensor modeling of room impulse responses,''
\newblock in {\em Proc. European Signal Process. Conf. ({EUSIPCO})}, Aug. 2021,
  pp. 111--115.

\bibitem{helwaniGenerative2023}
K.~Helwani, P.~Smaragdis, and M.~M. Goodwin,
\newblock ``Generative modeling based manifold learning for adaptive filtering
  guidance,''
\newblock in {\em Proc. {IEEE} Int. Conf. Acoust., Speech, Signal Process.
  ({ICASSP})}, 2023.

\bibitem{elkoRoom2003}
G.~Elko, E.~Diethorn, and T.~Gaensler,
\newblock ``Room impulse response variation due to thermal fluctuation and its
  impact on acoustic echo cancellation,''
\newblock {\em Proc. Int. Workshop Acoust. Echo Noise Cont. ({IWAENC})}, Jan.
  2003.

\bibitem{prawdaTime2023}
K.~Prawda, S.~Schlecht, and V.~V{\"a}lim{\"a}ki,
\newblock ``Time variance in measured room impulse responses,''
\newblock in {\em Proc. Forum Acusticum}, Sept. 2023.

\bibitem{olsenSound2017}
M.~Olsen and M.~B. M{\o}ller,
\newblock ``Sound zones: On the effect of ambient temperature variations in
  feed-forward systems,''
\newblock in {\em Proc. {AES} Conv.} May 2017, pp. 1009--1018.

\bibitem{prawdaShorttime2024}
K.~Prawda, S.~J. Schlecht, and V.~V{\"a}lim{\"a}ki,
\newblock ``Short-time coherence between repeated room impulse response
  measurements,''
\newblock {\em J. Acoust. Soc. Am.}, vol. 156, no. 2, pp. 1017--1028, Aug.
  2024.

\bibitem{verburgReconstruction2018}
S.~A. Verburg and E.~{Fernandez-Grande},
\newblock ``Reconstruction of the sound field in a room using compressive
  sensing,''
\newblock {\em J. Acoust. Soc. Am.}, vol. 143, no. 6, pp. 3770--3779, June
  2018.

\bibitem{katzbergCompressed2018}
F.~Katzberg, R.~Mazur, M.~Maass, P.~Koch, and A.~Mertins,
\newblock ``A compressed sensing framework for dynamic sound-field
  measurements,''
\newblock {\em {IEEE/ACM} Trans. Audio, Speech, Lang. Process.}, vol. 26, no.
  11, pp. 1962--1975, Nov. 2018.

\bibitem{koyamaSparse2019}
S.~Koyama and L.~Daudet,
\newblock ``Sparse representation of a spatial sound field in a reverberant
  environment,''
\newblock {\em IEEE J. Sel. Top. Signal Process.}, vol. 13, no. 1, pp.
  172--184, Mar. 2019.

\bibitem{damianoCompressive2024}
S.~Damiano, F.~Borra, A.~Bernardini, F.~Antonacci, and A.~Sarti,
\newblock ``A compressive sensing approach for the reconstruction of the
  soundfield produced by directive sources in reverberant rooms,''
\newblock {\em {IEEE/ACM} Trans. Audio, Speech, Lang. Process.}, vol. 32, pp.
  2667--2679, 2024.

\bibitem{caviedes-nozalSpatiotemporal2023}
D.~{Caviedes-Nozal} and E.~{Fernandez-Grande},
\newblock ``Spatio-temporal bayesian regression for room impulse response
  reconstruction with spherical waves,''
\newblock {\em {IEEE/ACM} Trans. Audio, Speech, Lang. Process.}, vol. 31, pp.
  3263--3277, 2023.

\bibitem{schmidSpatial2021}
J.~M. Schmid, E.~{Fernandez-Grande}, M.~Hahmann, C.~Gurbuz, M.~Eser, and
  S.~Marburg,
\newblock ``Spatial reconstruction of the sound field in a room in the modal
  frequency range using {{Bayesian}} inference,''
\newblock {\em J. Acoust. Soc. Am.}, vol. 150, no. 6, pp. 4385--4394, Dec.
  2021.

\bibitem{karakonstantisGenerative2023}
X.~Karakonstantis and E.~{Fernandez-Grande},
\newblock ``Generative adversarial networks with physical sound field priors,''
\newblock {\em J. Acoust. Soc. Am.}, vol. 154, no. 2, pp. 1226--1238, Aug.
  2023.

\bibitem{miotelloDeep2024}
F.~Miotello, M.~Pezzoli, L.~Comanducci, F.~Antonacci, and A.~Sarti,
\newblock ``Deep prior-based audio inpainting using multi-resolution harmonic
  convolutional neural networks,''
\newblock {\em {IEEE/ACM} Trans. Audio, Speech, Lang. Process.}, vol. 32, pp.
  113--123, 2024.

\bibitem{shigemiPhysicsinformed2022}
K.~Shigemi, S.~Koyama, T.~Nakamura, and H.~Saruwatari,
\newblock ``Physics-informed convolutional neural network with bicubic spline
  interpolation for sound field estimation,''
\newblock in {\em Proc. Int. Workshop Acoust. Signal Enhancement ({IWAENC})},
  Sept. 2022.

\bibitem{lluisSound2020}
F.~Llu{\'i}s, P.~{Mart{\'i}nez-Nuevo}, M.~Bo~M{\o}ller, and S.~Ewan~Shepstone,
\newblock ``Sound field reconstruction in rooms: {{Inpainting}} meets
  super-resolution,''
\newblock {\em J. Acoust. Soc. Am.}, vol. 148, no. 2, pp. 649--659, Aug. 2020.

\bibitem{olivieriPhysicsinformed2024}
M.~Olivieri, X.~Karakonstantis, M.~Pezzoli, F.~Antonacci, A.~Sarti, and
  E.~{Fernandez-Grande},
\newblock ``Physics-informed neural network for volumetric sound field
  reconstruction of speech signals,''
\newblock {\em EURASIP Journal on Audio, Speech, and Music Processing}, vol.
  2024, no. 1, pp. 42, Sept. 2024.

\bibitem{antonelloRoom2017}
N.~Antonello, E.~D. Sena, M.~Moonen, P.~A. Naylor, and T.~van Waterschoot,
\newblock ``Room impulse response interpolation using a sparse spatio-temporal
  representation of the sound field,''
\newblock {\em {IEEE/ACM} Trans. Audio, Speech, Lang. Process.}, vol. 25, no.
  10, pp. 1929--1941, Oct. 2017.

\bibitem{sundstromOptimal2024}
D.~Sundstr{\"o}m, F.~Elvander, and A.~Jakobsson,
\newblock ``Optimal transport based impulse response interpolation in the
  presence of calibration errors,''
\newblock {\em IEEE Trans. Signal Process.}, vol. 72, pp. 1548--1559, 2024.

\bibitem{saburouTheory2016}
S.~Saburou and Y.~Sawano,
\newblock {\em Theory of Reproducing Kernels and Applications}, vol.~44 of {\em
  Developments in Mathematics},
\newblock Springer, 2016.

\bibitem{uenoKernel2018}
N.~Ueno, S.~Koyama, and H.~Saruwatari,
\newblock ``Kernel ridge regression with constraint of {{Helmholtz}} equation
  for sound field interpolation,''
\newblock in {\em Proc. Int. Workshop Acoust. Signal Enhancement ({IWAENC})}.
  Sept. 2018, pp. 436--440.

\bibitem{itoFeedforward2019}
H.~Ito, S.~Koyama, N.~Ueno, and H.~Saruwatari,
\newblock ``Feedforward spatial active noise control based on kernel
  interpolation of sound field,''
\newblock in {\em Proc. {IEEE} Int. Conf. Acoust., Speech, Signal Process.
  ({ICASSP})}. May 2019, pp. 511--515.

\bibitem{brunnstromKernelinterpolationbased2021}
J.~Brunnstr{\"o}m and S.~Koyama,
\newblock ``Kernel-interpolation-based filtered-x least mean square for spatial
  active noise control in time domain,''
\newblock in {\em Proc. {IEEE} Int. Conf. Acoust., Speech, Signal Process.
  ({ICASSP})}, June 2021, pp. 161--165.

\bibitem{koyamaSpatial2021}
S.~Koyama, J.~Brunnstr{\"o}m, H.~Ito, N.~Ueno, and H.~Saruwatari,
\newblock ``Spatial active noise control based on kernel interpolation of sound
  field,''
\newblock {\em {IEEE/ACM} Trans. Audio, Speech, Lang. Process.}, vol. 29, pp.
  3052--3063, Aug. 2021.

\bibitem{brunnstromVariable2022}
J.~Brunnstr{\"o}m, S.~Koyama, and M.~Moonen,
\newblock ``Variable span trade-off filter for sound zone control with kernel
  interpolation weighting,''
\newblock in {\em Proc. {IEEE} Int. Conf. Acoust., Speech, Signal Process.
  ({ICASSP})}, May 2022, pp. 1071--1075.

\bibitem{koyamaWeighted2022}
S.~Koyama and K.~Arikawa,
\newblock ``Weighted pressure matching based on kernel interpolation for sound
  field reproduction,''
\newblock in {\em Proc. Int. Congr. Acoust. ({ICA})}, Oct. 2022.

\bibitem{uenoDirectionally2021}
N.~Ueno, S.~Koyama, and H.~Saruwatari,
\newblock ``Directionally weighted wave field estimation exploiting prior
  information on source direction,''
\newblock {\em IEEE Trans. Signal Process.}, vol. 69, pp. 2383--2395, Apr.
  2021.

\bibitem{ribeiroSound2024}
J.~Ribeiro, S.~Koyama, and H.~Saruwatari,
\newblock ``Sound field estimation based on physics-constrained kernel
  interpolation adapted to environment,''
\newblock {\em {IEEE/ACM} Trans. Audio, Speech, Lang. Process.}, vol. 32, pp.
  4369--4383, Sept. 2024.

\bibitem{samarasingheWavefield2014}
P.~Samarasinghe, T.~Abhayapala, and M.~Poletti,
\newblock ``Wavefield analysis over large areas using distributed higher order
  microphones,''
\newblock {\em {IEEE/ACM} Trans. Audio, Speech, Lang. Process.}, vol. 22, no.
  3, pp. 647--658, Mar. 2014.

\bibitem{samarasingheEfficient2015}
P.~Samarasinghe, T.~Abhayapala, M.~Poletti, and T.~Betlehem,
\newblock ``An efficient parameterization of the room transfer function,''
\newblock {\em {IEEE/ACM} Trans. Audio, Speech, Lang. Process.}, vol. 23, no.
  12, pp. 2217--2227, Dec. 2015.

\bibitem{uenoSound2018}
N.~Ueno, S.~Koyama, and H.~Saruwatari,
\newblock ``Sound field recording using distributed microphones based on
  harmonic analysis of infinite order,''
\newblock {\em IEEE Signal Process. Lett.}, vol. 25, no. 1, pp. 135--139, Jan.
  2018.

\bibitem{brunnstromBayesian2024}
J.~Brunnstr{\"o}m, M.~B. M{\o}ller, J.~{\O}stergaard, and M.~Moonen,
\newblock ``Bayesian sound field estimation using uncertain data,''
\newblock in {\em Proc. Int. Workshop Acoust. Signal Enhancement ({IWAENC})},
  Sept. 2024, pp. 329--333.

\bibitem{brunnstromBayesian2025}
J.~Brunnstr{\"o}m, M.~B. M{\o}ller, and M.~Moonen,
\newblock ``Bayesian sound field estimation using moving microphones,''
\newblock {\em IEEE Open J. Signal Process.}, vol. 6, pp. 312--322, Jan. 2025.

\bibitem{rasmussenGaussian2006}
C.~E. Rasmussen and C.~K.~I. Williams,
\newblock {\em Gaussian Processes for Machine Learning},
\newblock Adaptive Computation and Machine Learning. MIT Press, 2006.

\bibitem{caviedes-nozalGaussian2021}
D.~{Caviedes-Nozal}, N.~A.~B. Riis, F.~M. Heuchel, J.~Brunskog, P.~Gerstoft,
  and E.~{Fernandez-Grande},
\newblock ``Gaussian processes for sound field reconstruction,''
\newblock {\em J. Acoust. Soc. Am.}, vol. 149, no. 2, pp. 1107--1119, Feb.
  2021.

\bibitem{fernandez-grandeReconstruction2021}
E.~{Fernandez-Grande}, D.~{Caviedes-Nozal}, M.~Hahmann, X.~Karakonstantis, and
  S.~A. Verburg,
\newblock ``Reconstruction of room impulse responses over extended domains for
  navigable sound field reproduction,''
\newblock in {\em Proc. Int. Conf. Immersive 3D Audio}, Sept. 2021.

\bibitem{fengRoom2024}
X.~Feng, J.~Cheng, S.~Chen, and Y.~Shen,
\newblock ``Room impulse response reconstruction using pattern-coupled sparse
  bayesian learning with spherical waves,''
\newblock {\em IEEE Signal Process. Lett.}, vol. 31, pp. 1925--1929, 2024.

\bibitem{liangSound2024}
Z.~Liang, W.~Zhang, and T.~D. Abhayapala,
\newblock ``Sound field reconstruction using neural processes with dynamic
  kernels,''
\newblock {\em EURASIP J. Audio Speech Music Process.}, vol. 2024, Feb. 2024.

\bibitem{scholkopfGeneralized2001}
B.~Sch{\"o}lkopf, R.~Herbrich, and A.~J. Smola,
\newblock ``{A generalized representer theorem},''
\newblock in {\em Proc. Int. Conf. Comput. Learn. Theory}. July 2001, pp.
  416--426.

\bibitem{dinuzzoRepresenter2012}
F.~Dinuzzo and B.~Sch{\"o}lkopf,
\newblock ``The representer theorem for {{Hilbert}} spaces: A necessary and
  sufficient condition,''
\newblock in {\em Adv. {{Neural Inf}}. {{Process}}. {{Syst}}.} 2012, vol.~25.

\bibitem{argyriouUnifying2014}
A.~Argyriou and F.~Dinuzzo,
\newblock ``A unifying view of representer theorems,''
\newblock in {\em Proc. Int. Conf. Mach. Learn. (ICML)}, June 2014, vol.~32,
  pp. 748--756.

\bibitem{minhUnifying2016}
H.~Q. Minh, L.~Bazzani, and V.~Murino,
\newblock ``A unifying framework in vector-valued reproducing kernel hilbert
  spaces for manifold regularization and co-regularized multi-view learning,''
\newblock {\em J. Mach. Learn. Res.}, vol. 17, pp. 1--72, 2016.

\bibitem{diwaleGeneralized2018}
S.~Diwale and C.~Jones,
\newblock ``A generalized representer theorem for {{Hilbert}} space-valued
  functions,'' Sept. 2018,
\newblock arXiv preprint, arXiv:1809.07347.

\bibitem{boyerRepresenter2019}
C.~Boyer, A.~Chambolle, Y.~D. Castro, V.~Duval, F.~{de Gournay}, and P.~Weiss,
\newblock ``On representer theorems and convex regularization,''
\newblock {\em SIAM J. Optim.}, vol. 29, no. 2, pp. 1260--1281, Jan. 2019.

\bibitem{sundstromSound2024}
D.~Sundstr{\"o}m, S.~Koyama, and A.~Jakobsson,
\newblock ``Sound field estimation using deep kernel learning regularized by
  the wave equation,''
\newblock in {\em Proc. Int. Workshop Acoust. Signal Enhancement ({IWAENC})},
  Sept. 2024, pp. 319--323.

\bibitem{martinMultiple2006}
P.~A. Martin,
\newblock {\em Multiple Scattering: {{Interaction}} of Time-Harmonic Waves with
  {{N}} Obstacles}, vol. 107 of {\em Encyclopedia of Mathematics and Its
  Applications},
\newblock Cambridge University Press, 2006.

\bibitem{bradburyJAX2018}
J.~Bradbury, R.~Frostig, P.~Hawkins, M.~J. Johnson, C.~Leary, D.~Maclaurin,
  G.~Necula, A.~Paszke, J.~VanderPlas, S.~{Wanderman-Milne}, and Q.~Zhang,
\newblock ``{{JAX}}: {{Composable}} transformations of {{Python}}+{{NumPy}}
  programs,'' 2018,
\newblock [Online]. Available: http://github.com/jax-ml/jax.

\bibitem{harrisArray2020}
C.~R. Harris, K.~J. Millman, S.~J. {van der Walt}, R.~Gommers, P.~Virtanen,
  D.~Cournapeau, E.~Wieser, J.~Taylor, S.~Berg, N.~J. Smith, R.~Kern, M.~Picus,
  S.~Hoyer, M.~H. {van Kerkwijk}, M.~Brett, A.~Haldane, J.~F. {del R{\'i}o},
  M.~Wiebe, P.~Peterson, P.~{G{\'e}rard-Marchant}, K.~Sheppard, T.~Reddy,
  W.~Weckesser, H.~Abbasi, C.~Gohlke, and T.~E. Oliphant,
\newblock ``Array programming with {{NumPy}},''
\newblock {\em Nature}, vol. 585, no. 7825, pp. 357--362, Sept. 2020.

\bibitem{kennedyHilbert2013}
R.~A. Kennedy and P.~Sadeghi,
\newblock {\em Hilbert Space Methods in Signal Processing},
\newblock Cambridge University Press, Mar. 2013.

\bibitem{williamsFourier1999}
E.~G. Williams and J.~A. Mann,
\newblock {\em Fourier Acoustics: Sound Radiation and Nearfield Acoustical
  Holography}, vol. 108,
\newblock Academic Press, 1999.

\bibitem{ribeiroPhysicsconstrained2024}
J.~G.~C. Ribeiro, S.~Koyama, and H.~Saruwatari,
\newblock ``Physics-constrained adaptive kernel interpolation for
  region-to-region acoustic transfer function: A {{Bayesian}} approach,''
\newblock {\em EURASIP Journal on Audio, Speech, and Music Processing}, vol.
  2024, no. 1, pp. 43, Sept. 2024.

\bibitem{mardiaDirectional2000}
K.~Mardia and P.~Jupp,
\newblock {\em Directional Statistics},
\newblock Wiley Series in Probability and Statistics. John Wiley \& Sons, 2000.

\bibitem{vanwaterschootOptimal2005}
T.~{van Waterschoot}, G.~Rombouts, and M.~Moonen,
\newblock ``Towards optimal regularization by incorporating prior knowledge in
  an acoustic echo canceller,''
\newblock in {\em Proc. Int. Workshop Acoust. Echo Noise Cont. ({IWAENC})},
  Sept. 2005, pp. 157--160.

\bibitem{allenImage1979}
J.~B. Allen and D.~A. Berkley,
\newblock ``Image method for efficiently simulating small-room acoustics,''
\newblock {\em J. Acoust. Soc. Am.}, vol. 65, no. 4, pp. 943--950, Apr. 1979.

\bibitem{scheiblerPyroomacoustics2018}
R.~Scheibler, E.~Bezzam, and I.~Dokmani{\'c},
\newblock ``Pyroomacoustics: {{A Python}} package for audio room simulations
  and array processing algorithms,''
\newblock in {\em Proc. {IEEE} Int. Conf. Acoust., Speech, Signal Process.
  ({ICASSP})}, Apr. 2018, pp. 351--355.

\bibitem{koyamaMeshRIR2021}
S.~Koyama, T.~Nishida, K.~Kimura, T.~Abe, N.~Ueno, and J.~Brunnstr{\"o}m,
\newblock ``{{MeshRIR}}: A dataset of room impulse responses on meshed grid
  points for evaluating sound field analysis and synthesis methods,''
\newblock in {\em Proc. {IEEE} Int. Workshop Appl. Signal Process. Audio
  Acoust. ({WASPAA})}, Oct. 2021, pp. 151--155.

\bibitem{antweilerPerfect1994}
C.~Antweiler and M.~D{\"o}rbecker,
\newblock ``Perfect sequence excitation of the {{NLMS}} algorithm and its
  application to acoustic echo control,''
\newblock {\em Ann. T{\'e}l{\'e}commun.}, vol. 49, no. 7, pp. 386--397, July
  1994.

\bibitem{antweilerNLMStype2008}
C.~Antweiler, A.~Telle, and P.~Vary,
\newblock ``{{NLMS-type}} system identification of {{MISO}} systems with
  shifted perfect sequences,''
\newblock in {\em Proc. Int. Workshop Acoust. Signal Enhancement ({IWAENC})},
  Sept. 2008.

\bibitem{nelkeMeasurement2014}
C.~M. Nelke and P.~Vary,
\newblock ``Measurement, analysis and simulation of wind noise signals for
  mobile communication devices,''
\newblock in {\em Proc. Int. Workshop Acoust. Signal Enhancement ({IWAENC})},
  Sept. 2014, pp. 327--331.

\bibitem{nelkeWind2014}
C.~Nelke,
\newblock ``Wind noise database,'' 2014,
\newblock [Online]. Available:
  https://www.iks.rwth-aachen.de/forschung/tools-downloads/databases/wind-noise-database.

\bibitem{conwayCourse1990}
J.~B. Conway,
\newblock {\em A Course in Functional Analysis},
\newblock Graduate Texts in Mathematics. Springer, 2nd edition, Sept. 1990.

\end{thebibliography}
	
\end{document}